\documentclass[a4paper]{article}
\usepackage{amsfonts,amssymb,amsmath,amsthm,cite}
\usepackage{ucs} 
\usepackage[utf8x]{inputenc}
\usepackage{graphicx}
\usepackage[english]{babel}
\usepackage{slashed}
\usepackage{textcomp}
\usepackage{bm}

\evensidemargin=\oddsidemargin
\begin{document}
\begin{center}
	{\Large\textbf{Formula for two-loop divergent part\\ of 4-D Yang--Mills effective action}}
	\vspace{0.5cm}
	
	{\large A.~V.~Ivanov$^\dag$ and N.~V.~Kharuk$^\ddag$}
	
	\vspace{0.5cm}
	
	$^\dag$$^\ddag${\it St. Petersburg Department of Steklov Mathematical Institute of
		Russian Academy of Sciences,}\\{\it 27 Fontanka, St. Petersburg 191023, Russia}\\
	$^\dag$$^\ddag${\it Leonhard Euler International Mathematical Institute, 10 Pesochnaya nab.,}\\
	{\it St. Petersburg 197022, Russia}\\
	$^\dag${\it E-mail: regul1@mail.ru}~~~~$^\ddag${\it E-mail: natakharuk@mail.ru}	
\end{center}
\vskip 10mm
\date{\vskip 20mm}
\vskip 10mm
\begin{abstract}
In the paper, we study the two-loop contribution to the effective action of the four-dimensional quantum Yang--Mills theory. We derive a new formula for the contribution in terms of three functions, formed from the Green's function expansion near the diagonal. This result can be applied to different types of regularization. Therefore, we test it by using the dimensional regularization and cutoff ones and show the consistence with the results, obtained in other works.
\end{abstract}
\normalsize
\tableofcontents

\newpage
\section{Introduction}
\label{sec:intro}
The Yang--Mills fields firstly appeared in the paper \cite{1}. These objects have quite natural geometrical \cite{Trautman,Babelon,2} and physical \cite{8} interpretations that leads to their fundamental nature and relevance in the modern theoretical and mathematical physics.
The quantum theory of these fields has a number of mathematical problems nowadays. Let us consider one of them. 

As it is known, the most popular tool to investigate the Yang--Mills theory is the perturbative expansion (with the use of the Feynman diagrams \cite{4}) of the path integral, see \cite{3}. Such way is quite fruitful, but every term of the decomposition can contain integrals that do not converge and, hence, should be regularized. In this case we need to use the renormalization theory \cite{6,7,105} that makes the Yang--Mills theory physically meaningful and finite. At the same time the use of the renormalization procedure depends on the type of regularization \cite{9,10}.

One of the most common types of regularization are dimensional \cite{5,19} and cutoff \cite{16,17,34,Ivanov-Kharuk-2020}. Each approach has its own pros and cons. For example, the dimensional regularization allows simple version of multi-loop calculations \cite{11,12,13,14,15,18,20,Ivanov-2019} and preserves a gauge invariance. However, it does not have a physical nature, because we need to work in non-integer-dimensional space. Another example is the cutoff regularization that has quite clear physical nature, but it can violate the gauge invariance and allows the appearance of non-logarithmic divergences, see \cite{w5,w6,w7,w8}. Of course, there are other types of regularization, such as Pauli--Villars \cite{Pauli-Villars} or regularization by higher covariant derivatives \cite{3,Bakeyev-Slavnov}, but they are not considered in the paper.


In the present work we study an infrared part of the two-loop contribution to the Yang--Mills effective action. We derive a new formula for this part in terms of three functions, which follow from the expansion of the Green's function near the diagonal. At the same time we do not concretize the scheme of the regularization, so the formula has general nature. As an example, we test our formula using different popular types of regularization and demonstrate consistency of the results.

We believe that our results are useful and interesting, because they give the ability to investigate regularizations on the example of the four-dimensional Yang--Mills theory. As it is mentioned above, not any regularization satisfies all required properties. Hence, this is very important and helpful to have a simple way to check and control.

The structure of the work is the following. In Section \ref{sec:bas} we introduce basic information, such as properties of the Yang--Mills theory and the heat kernel expansion,  and formulate the main results. Then, in Section \ref{sec:modver} we introduce new types of vertices for working with the perturbative expansion. After that, in Section \ref{sec:twol} we derive and prove the main result, and in Section \ref{sec:exam:} we test the final formula by using the dimensional and cutoff regularizations. In the conclusion we give a few remarks.
\section{Basic concepts and results}
\label{sec:bas}
\subsection{Yang--Mills theory}
\label{sec:bas:ym}
Let $G$ be a compact semisimple Lie group \cite{2}, and $\mathfrak{g}$ is its Lie algebra of a dimension $\dim\mathfrak{g}$.
Let $t^a$ be the generators of the algebra $\mathfrak{g}$, where $a=1,\ldots,\dim\mathfrak{g}$,
such that the relations hold
\begin{equation}\label{constdef}
[t^a,t^b]=f^{abc}t^c,\,\,\,\,\,\,\mathrm{tr}(t^at^b)=-2\delta^{ab},
\end{equation}
where $f^{abc}$ are antisymmetric structure constants for $\mathfrak{g}$, and '$\mathrm{tr}$' is the Killing form. We work with an adjoint representation, so it is easy to verify that the structure constants have the following crucial properties
\begin{equation}\label{constprop}
f^{abc}f^{aef}=f^{abf}f^{aec}-f^{acf}f^{aeb},\,\,\,
f^{abc}f^{abe}=c_2\delta^{ce}.
\end{equation}

Let $x,y\in U$, where $U$ is a smooth convex open domain from $\mathbb{R}^d$, and Greek letters $\mu,\nu$ denote the coordinate components. Then, by symbol $B^{\phantom{a}}_\mu(x)=B^a_\mu(x)t^a$, where $B^{\phantom{a}}_\mu(\cdot)\in C^{\infty}(U,\mathfrak{g})$ for all values of $\mu$, we define the components of a Yang--Mills connection. The operator $B^{\phantom{a}}_\mu(x)$ as an element of the Lie algebra acts by commutator according to the adjoint representation. Hence, we treat $B^{\phantom{a}}_\mu(x)$ as a matrix-valued operator with the components $f^{adb}B^{d}_\mu(x)$.

Then, after introducing the components of the field strength tensor in the form
\begin{equation*}
F_{\mu\nu}^a=\partial_\mu^{}B_\nu^a-\partial_\nu^{}B_\mu^a+f^{abc}B_\mu^bB_\nu^c,
\end{equation*}
we can formulate a classical action of the Yang--Mills theory \cite{3}
\begin{equation}\label{Clact}
S[B]=\frac{1}{4g^2}\int_{\mathbb{R}^4}d^4x\,F^a_{\mu\nu}F^{a}_{\mu\nu}=
\frac{W_{-1}}{4g^2},
\end{equation} 
where $g$ is a coupling constant, and $W_{-1}=W_{-1}[B]$ is an auxiliary functional \cite{21,22,Ivanov-2018}.

Further, we are going to present a formula for a pure effective action. For the purpose, we need to introduce several additional objects. First of all we define the left and the right derivatives. Let $h(\cdot)\in C^1(U,\mathfrak{g})$ be an operator, and $h^{ab}(x)$ be its matrix components in the point $x$, then
\begin{equation}\label{deffs}
\overrightarrow{D}^{ab}_{x^{\mu}}h^{bc}(x)=\partial_{x^{\mu}}^{\phantom{a}}h^{ac}(x)+f^{adb}B^d_\mu(x)h^{bc}(x),\,\,\,
h^{ab}(x)\overleftarrow{D}^{bc}_{x^{\mu}}=\partial_{x^{\mu}}^{\phantom{a}}h^{ac}(x)-h^{ab}(x)f^{bdc}B^d_\mu(x).
\end{equation}

Next we give formulae for auxiliary differential operators
\begin{equation}\label{oper}
M_0^{ab}=-\overrightarrow{D}_\mu^{ae}\overrightarrow{D}_\mu^{eb},\,\,\,
M_{1\mu\nu}^{\,\,\,ab}=M_0^{ab}\delta_{\mu\nu}^{}-2f^{acb}F_{\mu\nu}^c,
\end{equation}
and vertex operators with functional derivatives
\begin{equation}\label{versh1}
\Gamma_1=-\int_{\mathbb{R}^4}d^4x\,\frac{\delta}{\delta J_\nu^{\,a}}\overrightarrow{D}_\mu^{ab}F_{\mu\nu}^b,\,\,\,\,\,\,
\Gamma_3=\int_{\mathbb{R}^4}d^4x\,\bigg(\overrightarrow{D}_\mu^{ae}\frac{\delta}{\delta J_\nu^{\,e}}\bigg)f^{abc}\frac{\delta}{\delta J_\mu^{\,b}}\frac{\delta}{\delta J_\nu^{\,c}},
\end{equation}
\begin{equation}\label{versh2}
\Gamma_4=\frac{1}{4}\int_{\mathbb{R}^4}d^4x\,f^{abc}\frac{\delta}{\delta J_\mu^{\,b}}\frac{\delta}{\delta J_\nu^{\,c}}f^{aed}\frac{\delta}{\delta J_\mu^{\,e}}\frac{\delta}{\delta J_\nu^{\,d}},\,\,\,\,\,\,
\Omega_3=\int_{\mathbb{R}^4}d^4x\,\bigg(\overrightarrow{D}_\mu^{ab}\frac{\delta}{\delta b^{\,b}}\bigg)f^{aed}\frac{\delta}{\delta J_\mu^{\,e}}\frac{\delta}{\delta\bar{b}^{\,d}},
\end{equation}
where $J_\mu^{\,a}$ and the ghost fields $b^{\,a}$ and $\bar{b}^{\,a}$, see \cite{27}, have smooth densities. Then let us define the Green's functions $G_0$ and $G_1$ for the Laplace-type operators $M_0$ and $M_1$ by the equalities
\begin{equation}
\label{green}
M_{1\mu\nu}^{\,\,\,ab}G_{1\nu\rho}^{\,\,\,bc}(x,y)=\delta^{ac}\delta_{\mu\rho}\delta(x-y),\,\,\,\,\,\,
M_0^{ab}G_{0}^{bc}(x,y)=\delta^{ac}\delta(x-y).
\end{equation}

\begin{figure}[h]
	\centerline{\includegraphics[width=0.5\linewidth]{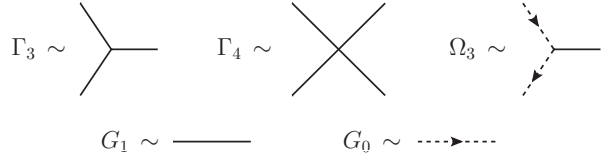}}
	\caption{Diagram technique elements.}
	\label{loop1}
\end{figure}

We note that according to the rules of Feynman diagram technique, formulae (\ref{versh1}), (\ref{versh2}), and (\ref{green}) are connected to their diagrammatic representation, see  \cite{104,23} and Figure \ref{loop1}.

Now we are ready to introduce a pure effective action for the Yang--Mills theory. Let us
apply the background field method \cite{24,25,26,101,102,103} to the path integral formulation of the Yang--Mills theory. Also, we define an additional functional of $B^a_\mu$
\begin{equation}
\label{eq20}
W[B]=S[B]+\bigg\{\frac{1}{2}\ln\det(M_1/M_1|_{B=0})-\ln\det(M_0/M_0|_{B=0})\bigg\}+
W_{h}[B],
\end{equation}
where a contribution $W_{h}[B]$ for higher loops has the following form
\begin{equation}
\label{highloops}
W_{h}[B]=-\ln\bigg(\exp\big(-\Gamma_1/g-g\Gamma_3-g^2\Gamma_4+g\,\Omega_3\big)\,
Z[J,b,\bar{b}\,]\Big|_{J_\mu=b=\bar{b}=0} \bigg)\bigg|_{\mathrm{1PI}\,\,\mathrm{part}},
\end{equation}
and the generating functional $Z[J,b,\bar{b}]=\exp(g_1+g_0)$ consists of
\begin{equation}\label{oldgreen}
g_1=
\frac{1}{2}\int_{\mathbb{R}^4}d^4x\int_{\mathbb{R}^4}d^4y\,J_\mu^a(x) G_{1\mu\nu}^{\,\,\,ab}(x,y)\,J_\nu^b(y),\,\,\,
g_0=\int_{\mathbb{R}^4}d^4x\int_{\mathbb{R}^4}d^4y\,\bar{b}^{\,a}(x)G_0^{ab}(x,y)b^{\,b}(y).
\end{equation}

Then the \textbf{pure effective action} can be represented in the following form
\begin{equation}
\label{eq46}
W_{\mathrm{eff}}[B]=W[B]-W[0].
\end{equation}

\subsection{Heat kernel expansion}
\label{sec:bas:hk}
The main object in the heat kernel expansion is a path-ordered exponential. Let us give an appropriate definition by the following formula
\begin{equation}\label{expdef}
\Phi^{ab}(x,y)=\delta^{ab}+\sum_{k=1}^{+\infty}(-1)^k
\int_0^1ds_1\ldots\int_0^{s_{k-1}}ds_k\,
(x-y)^{\mu_1\ldots\mu_k}\Big(f^{ad_1c_1}B^{d_1}_{\mu_1}(z(s_1))\Big)\ldots
\Big(f^{c_{k-1}d_kb}B^{d_k}_{\mu_1}(z(s_k))\Big),
\end{equation}
where $z^\mu(s)=y^\mu+s(x-y)^\mu$, see \cite{Shore,Polyakov}. 

Such type of operators has some useful properties, that can be formulated in the form
\begin{equation}\label{expprop}
\Phi^{ab}(x,z)\Phi^{bc}(z,y)=\Phi^{ac}(x,y),\,\,
(\Phi^{-1})^{ab}(x,y)=\Phi^{ab}(y,x)=\Phi^{ba}(x,y),\,\,
\Phi^{ab}(y,y)=\delta^{ab},
\end{equation}
where the point $z\in U$ belongs to a straight line passing through the points $x$ and $y$. In other words, it means that there is such $s\in\mathbb{R}$, that the equality $z^\mu=y^\mu+s(x-y)\in U$ holds. The proofs of the properties described above can be found in \cite{Shore,31,33}.

Therefore, we can formulate the differential equations for the exponential as
\begin{equation}\label{phieq}
(x-y)^\mu\overrightarrow{D}^{ab}_{x^{\mu}}\Phi^{bc}(x,y)=0\,\,\,\,\mbox{and}\,\,\,\,
\Phi^{ab}(x,y)\overleftarrow{D}^{bc}_{y^{\mu}}(x-y)^\mu=0.
\end{equation}
The proof can be achieved by straight differentiation of (\ref{expdef}) and integration by parts, see \cite{Shore,31}.

Now we want to remember some basic concepts of the heat kernel expansion and the corresponding useful results. Let us introduce a Laplace-type operator $A$, which has a more general view that in (\ref{oper}). Locally, it has the following form
\begin{equation}\label{oper1}
A^{ab}(x)=-IM_0^{cd}(x)-v^{ab}(x),
\end{equation}
where $I$ is an arbitrary $n\times n$ with $n\in\mathbb{N}$, and $v^{ab}(x)$ is a $n\times n$ matrix-valued smooth potential, such that the operator $A$ is symmetric. If we take $n=4$, $(I)_{\mu\nu}=\delta_{\mu\nu}$, and $(v^{ab})_{\mu\nu}(x)=2f^{acb}F^d_{\mu\nu}(x)$, then we obtain the operator $M_{1\mu\nu}^{\,\,\,ab}(x)$. Also, for the convenience we will not write the unit matrix $I$ in the rest of the text, because this does not create confusion.

Then from the general theory we know that an asymptotic expansion of a solution 
of the problem
\begin{equation}\label{HK1}
\big(\delta^{ac}\partial_\tau+A^{ac}(x)\big)K^{cb}(x,y;\tau)=0,\,\,\,K^{ab}(x,y;0)=\delta^{ab}\delta(x-y),
\end{equation}
for enough small values of the proper time $\tau\to+0$ can be found in the form \cite{28,29,30,31,vas1,vas2}
\begin{equation}\label{HK2}
K^{ab}(x,y;\tau)=(4\pi\tau)^{-2}e^{-|x-y|^2/4\tau}\sum_{k=0}^{+\infty}\tau^k\mathfrak{a}_k^{ab}(x,y).
\end{equation}

The coefficient $\mathfrak{a}^{ab}(x,y)$ of expansion (\ref{HK2}), Seeley--DeWitt coefficients, can be calculated recurrently, because they satisfy the following system of equations
\begin{equation}\label{SD1}
\mathfrak{a}_{0}^{ab}(x,y)=\Phi^{ab}(x,y),\,\,\,
\big(k+(x-y)^\sigma\overrightarrow{D}_{x^\sigma}^{ac}\big)
\mathfrak{a}_{k}^{cb}(x,y)=-A^{ac}(x)
\mathfrak{a}_{k-1}^{cb}(x,y),\,\,\,k\geqslant 1.
\end{equation}

The operators $M_0^{ab}$ and $M_{1\mu\nu}^{\,\,\,ab}$ for the Yang--Mills theory are special cases of the operator $A^{ab}$. Hence, using the formulae introduced above, we can write out the following asymptotic behaviour for the Green's function in the four-dimensional space 
\cite{29,psi}
\begin{multline}\label{SD9}
\big(A^{-1}\big)^{ab}(x,y)=R_0(x-y)\mathfrak{a}_{0}^{ab}(x,y)+
R_1(x-y)\mathfrak{a}_{1}^{ab}(x,y)\\+
R_2(x-y)\mathfrak{a}_{2}^{ab}(x,y)+\mathcal{PS}^{ab}(x,y)+\mathcal{ZM}^{ab}(x,y),
\end{multline}
where
\begin{equation}\label{SD10}
R_0(x)=\frac{1}{4\pi^2|x|^2},\,\,\,
R_1(x)=-\frac{\ln(|x|^2\mu^2)}{16\pi^2},\,\,\,
R_2(x)=\frac{|x|^2\big(\ln(|x|^2\mu^2)-1\big)}{64\pi^2},
\end{equation}
$\mathcal{PS}^{ab}$ is a non-local part, depending on the boundary conditions of a spectral problem, and $\mathcal{ZM}^{ab}$ is a number of local zero modes to satisfy the problem. Let us  note, it was shown in the paper \cite{Kharuk-2021}, that an infrared part in the second loop does not depend on $\mathcal{ZM}^{ab}$. Moreover, in the calculation process, we can choose $\mathcal{ZM}^{ab}$ in such a way, that the non-local part $\mathcal{PS}^{ab}$ would have the following behaviour near the diagonal $x\sim y$
\begin{equation}\label{SD11}
\mathcal{PS}^{ab}(x,y)=-\frac{|x-y|^2}{2^7\pi^2}\mathfrak{a}_{2}^{ab}(y,y)\big(1+o(1)\big).
\end{equation}

As it was noted in the papers \cite{Ivanov-Kharuk-2020,12,13}, the two-loop contribution to the $\beta$-function can contain only terms proportional to the classical action $W_{-1}$. This is beneficial observation, because we have the ability to consider a simplified version of the background field. The connection components have the form
\begin{equation}\label{singord3}
B^a_{\mu}(x)\to\tilde{B}^a_{\mu}(x)=\frac{1}{2}x^\nu \tilde{F}^a_{\nu\mu},
\end{equation}
where a new field strength $\big(\tilde{F}_{\nu\mu}\big)^{ac}=f^{abc}\tilde{F}^b_{\nu\mu}$ satisfies the following two equalities
\begin{equation}\label{singord4}
f^{acd}f^{deb}
\tilde{F}^{c}_{\nu\mu}\tilde{F}^{e}_{\sigma\rho}=
f^{acd}f^{deb}
\tilde{F}^{c}_{\sigma\rho}\tilde{F}^{e}_{\nu\mu}\,\,\,\,
\mbox{and}\,\,\,\,
\partial_{x^\sigma}\tilde{F}^a_{\nu\mu}=0\,\,\,\,\mbox{for all}\,\,\,\,
\mu,\nu,\sigma,\rho,a,b.
\end{equation}
The first relation means that the field strength is commutative (in the matrix sense), while the second one removes the dependence on all space variables. Additionally, we will require the normalization condition to be fulfilled $\tilde{F}_{\mu\nu}^a\tilde{F}_{\mu\nu}^a=1$. As an example, we can take the following matrix
\begin{equation}\label{singord6}
\big(\tilde{F}^a_{\mu\nu}\big)=\frac{1}{8\dim\mathfrak{g}}
\begin{pmatrix}
0&1&0&1\\
-1&0&1&0\\
0&-1&0&1\\
-1&0&-1&0
\end{pmatrix}\,\,\,
\mbox{for all}\,\,\,a\in\{1,\ldots,\dim\mathfrak{g}\}.
\end{equation}
\subsection{Results}
\label{sec:answtwoloop}
Now let us make some additional preparatory steps. First of all we should draw attention that we investigate the two-loop contribution to the effective action (\ref{eq46}). It means that we are interested in the terms from $W_h[B]-W_h[0]$ proportional to $g^2$, see formula (\ref{highloops}).

Let us define ten auxiliary constructions: $\mathrm{I}_9$ and $\mathrm{I}_{10}$ are from (\ref{aped9}), and eight integrals are defined by the following formulae
\begin{align}
\label{aped17}
\mathrm{I}_8&=c_2^2\int_{\mathrm{B}_{1/\mu}}d^dx\,
\Big(\partial_{x^\mu}R_0^{\phantom{2}}(x)\Big)
R_0^{\phantom{2}}(x)\partial_{x^\mu}
\bigg(\frac{|x|^2}{12d}R_1^{\phantom{2}}(x)+
\frac{1}{12}R_2^{\phantom{2}}(x)-
\frac{|x|^2}{2^93\pi^2}\bigg)
\bigg|^{\scriptsize{\mbox{IR-reg.}}},\\
\label{aped10}
\mathrm{I}_1&=c_2^2\int_{\mathrm{B}_{1/\mu}}d^dx\,
\Big(\partial_{x^\mu}R_0^{\phantom{2}}(x)\Big)
R_0^{\phantom{2}}(x)\partial_{x^\mu}
\bigg(\frac{|x|^2}{12d}R_1^{\phantom{2}}(x)+
\frac{(d-24)}{12d}R_2^{\phantom{2}}(x)+
\frac{(24-d)}{2^93d\pi^2}|x|^2\bigg)
\bigg|^{\scriptsize{\mbox{IR-reg.}}},
\\\label{aped11}
\mathrm{I}_2&=c_2^2\int_{\mathrm{B}_{1/\mu}}d^dx\,
\Big(\partial_{x^\mu}R_0^{\phantom{2}}(x)\Big)
\Big(\partial_{x^\mu}R_0^{\phantom{2}}(x)\Big)
\bigg(\frac{|x|^2}{12d}R_1^{\phantom{2}}(x)+
\frac{(d-24)}{12d}R_2^{\phantom{2}}(x)+
\frac{(24-d)}{2^93d\pi^2}|x|^2\bigg)
\bigg|^{\scriptsize{\mbox{IR-reg.}}}
,
\\\label{aped12}
\mathrm{I}_3&=-\frac{2c_2^2}{d}\int_{\mathrm{B}_{1/\mu}}d^dx\,
\Big(\partial_{x^\mu}R_0^{\phantom{2}}(x)\Big)
\Big(\partial_{x^\mu}R_1^{\phantom{2}}(x)\Big)R_1^{\phantom{2}}(x)
\Big|^{\scriptsize{\mbox{IR-reg.}}},
\\\label{aped13}
\mathrm{I}_4&=-2c_2^2\int_{\mathrm{B}_{1/\mu}}d^dx\,
R_0^{\phantom{2}}(x)\Big(\partial_{x^\mu}R_1^{\phantom{2}}(x)\Big)
\Big(\partial_{x^\mu}R_1^{\phantom{2}}(x)\Big)
\Big|^{\scriptsize{\mbox{IR-reg.}}},
\\
\label{aped14}
\mathrm{I}_5&=-\frac{c_2^2}{2d}\int_{\mathrm{B}_{1/\mu}}d^dx\,
R_1^{\phantom{2}}(x)x^\mu\partial_{x^\mu}R_0^{2}(x)\Big|^{\scriptsize{\mbox{IR-reg.}}},
\\\label{aped15}
\mathrm{I}_6&=\frac{c_2^2}{2d}\int_{\mathrm{B}_{1/\mu}}d^dx\,
R_0^{2}(x)x^\mu\partial_{x^\mu}R_1^{\phantom{2}}(x)\Big|^{\scriptsize{\mbox{IR-reg.}}},
\\\label{aped16}
\mathrm{I}_7&=\frac{c_2^2}{8d}\int_{\mathrm{B}_{1/\mu}}d^dx\,
|x|^2R_0^{3}(x)\Big|^{\scriptsize{\mbox{IR-reg.}}},
\end{align}
where the functions $R_0$, $R_1$, and $R_2$ were introduced in (\ref{SD10}), and the symbol "$\footnotesize{\mbox{IR-reg.}}$" shows that some type of infrared regularization has been applied. Additionally, the equal sign $\stackrel{\mathrm{IR}}{=}$ means that the constructions on both sides contain the same infrared logarithimic singularities. Non-logarithmic singularities, depending on the background field, do not appear in the calculations. At the same time all constants are cancelled due to definition (\ref{eq46}).

Let us formulate the main result of the paper. The divergent part of the multi-loop pure effective action, defined in formula (\ref{eq46}), has the following representation
\begin{equation}
\label{effact1}
W_h[B]-W_h[0]\Big|_{\footnotesize{\mbox{IR-reg.}}}\stackrel{\mathrm{IR}}{=}\eta W_{-1}
+o(g^2),
\end{equation}
where
\begin{multline}\label{twoloop112}
\eta\stackrel{\mathrm{IR}}{=}-\sum_{n=1}^6\mathcal{J}_n\stackrel{\mathrm{IR}}{=}
	g^2\bigg((3d-3)\mathrm{I}_1^{\phantom{0}}+\frac{(3d-4)}{2}\mathrm{I}_2^{\phantom{0}}+\frac{(d+2)}{2}\mathrm{I}_3^{\phantom{0}}
	+\frac{(2d-5)}{2d}\mathrm{I}_4^{\phantom{0}}\\
	+\frac{(8-d)}{2}\mathrm{I}_5^{\phantom{0}}+\frac{(d+2)}{2}
	\mathrm{I}_6^{\phantom{0}}+\frac{(3d-4)}{2}\mathrm{I}_7^{\phantom{0}}
	-\mathrm{I}_8^{\phantom{0}}+
	\frac{3}{2}\mathrm{I}_9^{\phantom{0}}+\frac{5}{2}\mathrm{I}_{10}^{\phantom{0}}
	\bigg).
\end{multline}

The simulations for four types of regularization, dimensional one and three types of cutoff one, are presented in Section \ref{sec:exam:cut}. All computations give proper results, consistent with the answers obtained earlier. Thereby, our new formula is confirmed and can be used in calculations with other different regularizations. We also compare the regularizations between themselves in Section \ref{sec:exam:} and show their pros and cons in the sense of computational difficulty.

\section{Modified vertices}
\label{sec:modver}
In the section we improve the diagram technique rules by introducing several types for each vertex. First of all, let us note that the standard vertices $\Gamma_3$ and $\Omega_3$ from (\ref{versh1}) and (\ref{versh2}) are linear functionals of the background field. Hence, we can divide them into two parts in the following way
\begin{equation}\label{simp1}
\Gamma_3^0=\int_{\mathbb{R}^d}d^dx\,\bigg(\partial_{x^{\mu}}\frac{\delta}{\delta J_\nu^{\,a}}\bigg)f^{abc}\frac{\delta}{\delta J_\mu^{\,b}}\frac{\delta}{\delta J_\nu^{\,c}},\,\,\,
\Gamma_3^1=\frac{1}{2}\int_{\mathbb{R}^d}d^dx\,\bigg(f^{ade}x^\sigma\tilde{F}_{\sigma\mu}^d\frac{\delta}{\delta J_\nu^{\,e}}\bigg)f^{abc}\frac{\delta}{\delta J_\mu^{\,b}}\frac{\delta}{\delta J_\nu^{\,c}},
\end{equation}
\begin{equation}\label{simp2}
\Omega_3^0=\int_{\mathbb{R}^d}d^dx\,\bigg(\partial_{x^{\mu}}\frac{\delta}{\delta b^{\,a}}\bigg)f^{abc}\frac{\delta}{\delta J_\mu^{\,b}}\frac{\delta}{\delta\bar{b}^{\,c}},
\,\,\,
\Omega_3^1=\frac{1}{2}\int_{\mathbb{R}^d}d^dx\,\bigg(f^{ade}x^\sigma\tilde{F}_{\sigma\mu}^d\frac{\delta}{\delta b^{\,a}}\bigg)f^{abc}\frac{\delta}{\delta J_\mu^{\,b}}\frac{\delta}{\delta\bar{b}^{\,c}},
\end{equation}
where we introduced the dimension of the space in a general way (by the symbol $d$), so that it would be possible to consider the dimensional regularization. Before the regularization is applied, it is equal to $4$.

\begin{figure}[h]
	\begin{minipage}[h]{0.48\linewidth}
		\center{\includegraphics[width=0.7\linewidth]{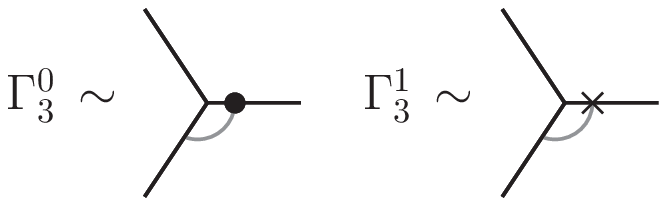}}
		\caption{Diagram technique elements for the new three-vertices (without the ghost field) defined in formula (\ref{simp1}).}
		\label{gamma}
	\end{minipage}
	\hfill
	\begin{minipage}[h]{0.48\linewidth}
		\center{\includegraphics[width=0.7\linewidth]{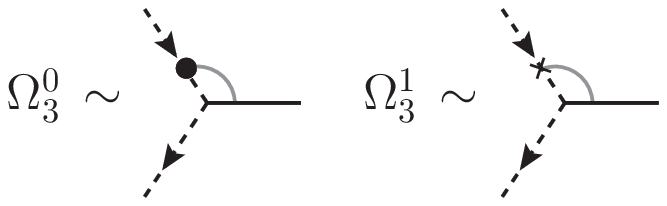}}
		\caption{Diagram technique elements for the new three-vertices (with the ghost field) defined in formula (\ref{simp2}).}
		\label{omega}
	\end{minipage}
\end{figure}

\begin{figure}[h]
	\begin{minipage}[h]{0.48\linewidth}
		\center{\includegraphics[width=0.63\linewidth]{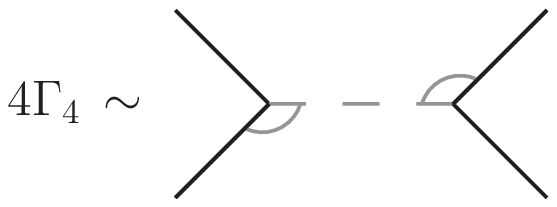}}
		\caption{Diagram technique element for the four-vertex.}
		\label{gamma4}
	\end{minipage}
	\hfill
	\begin{minipage}[h]{0.48\linewidth}
		\center{\includegraphics[width=0.77\linewidth]{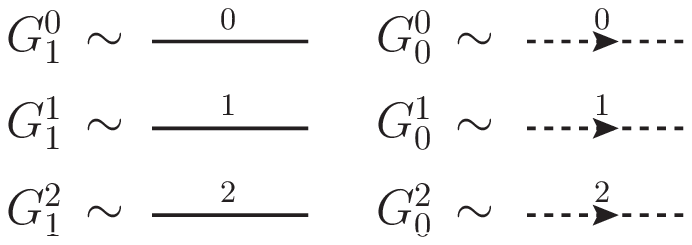}}
		\caption{Diagram technique elements for the new auxiliary functions, see formulae (\ref{newGreen5})--(\ref{newGreen7}).}\phantom{ddddddddddd}
		\label{propag}
	\end{minipage}
\end{figure}

According to the main idea we define the corresponding Feynman diagram technique for the new vertices. They are depicted in Figures \ref{gamma}--\ref{gamma4}, where we have marked the derivative $\partial_{x^\mu}$ by a black dot and the simplified background field $\tilde{B}_\mu$ by a cross. Such type of technique rules is a modified version of one suggested in the paper \cite{13}. Also, we should note that the arcs on the vertices symbolise the summation of the corresponding space indices, and the order of the external lines is related to the order of the group indices in the structure constant.

Also, we note that the new vertices and the previous ones satisfy the following relations
\begin{equation}\label{simp3}
\Gamma_3^0=\Gamma_3^{\phantom{0}}\big|_{B=0},\,\,\,
\Gamma_3^1=\Gamma_3^{\phantom{1}}\big|_{B\to\tilde{B}},\,\,\,
\Omega_3^0=\Omega_3^{\phantom{0}}\big|_{B=0},\,\,\,
\Omega_3^1=\Omega_3^{\phantom{1}}\big|_{B\to\tilde{B}}.
\end{equation}

To proceed we need to find the asymptotics for the initial Green's functions $G_0$ and $G_{1\,\nu\rho}$. They can be written as the series in powers of the background field components. For convenience, we define auxiliary functions $G_0^i$, $G^i_{1\,\nu\rho}$, where $i=0,1,2$. The functions have the following form
\begin{equation}\label{newGreen5}
G_0^1(x,y)=\frac{1}{2}x^\mu\tilde{F}_{\mu\sigma}y^\sigma
R_0(x-y)
,\,\,\,
G^1_{1\,\nu\rho}(x,y)=\delta_{\nu\rho}G_0^1(x,y)+2R_1(x-y)\tilde{F}_{\nu\rho},
\end{equation}
\begin{multline}\label{newGreen6}
G_0^2(x,y)=\frac{1}{4}\big(x^\mu\tilde{F}_{\mu\sigma}y^\sigma\big)^2
R_0(x-y)+\frac{1}{12}R_1(x-y)
(x-y)^{\alpha\beta}\tilde{F}_{\alpha\sigma}\tilde{F}_{\beta\sigma}\\+
\frac{1}{12}\bigg(R_2(x-y)-\frac{|x-y|^2}{2^7\pi^2}\bigg)
\tilde{F}_{\alpha\beta}\tilde{F}_{\alpha\beta},
\end{multline}
\begin{equation}\label{newGreen7}
G^2_{1\,\nu\rho}(x,y)=\delta_{\nu\rho}G_0^2(x,y)+R_2(x-y)
x^\mu\tilde{F}_{\mu\sigma}y^\sigma\tilde{F}_{\nu\rho}+
2\bigg(R_2(x-y)-\frac{|x-y|^2}{2^7\pi^2}\bigg)
\tilde{F}_{\nu\sigma}\tilde{F}_{\sigma\rho},
\end{equation}
where we have used definitions (\ref{SD10}).
Then, using the functions defined above and the results from the papers \cite{31,33,32,29}, we obtain the following decompositions for the Green's functions from (\ref{green}), when $s\to+0$,
\begin{equation}\label{newGreen1}
G_0^{\phantom{0}}(x,y)\Big|_{B\to s\tilde{B}}=G_0^0(x,y)+sG_0^1(x,y)+
s^2G_0^2(x,y)+\mathcal{O}(s^3),
\end{equation}
\begin{equation}\label{newGreen2}
G_{1\nu\rho}^{\phantom{0}}(x,y)\Big|_{B\to s\tilde{B}}=G_{1\nu\rho}^0(x,y)+sG_{1\nu\rho}^1(x,y)+
s^2G_{1\nu\rho}^2(x,y)+\mathcal{O}(s^3),
\end{equation}
where we have used an explicit formula for the path-ordered exponential (\ref{expdef}) in the particular case
\begin{equation}\label{newGreen8}
\Phi(x,y)\Big|_{B\to s\tilde{B}}=\exp\bigg(\frac{s}{2}x^\mu\tilde{F}_{\mu\sigma}y^\sigma\bigg).
\end{equation}

The diagram technique representation of the new functions is presented in Figure \ref{propag}, where the index symbolises the top index of the corresponding function.

Let us note that all new elements of the diagram technique have the top index, which symbolises the degree of the field strength tensor $\tilde{F}_{\mu\nu}$. This is quite convenient, because we can find a contribution, corresponding to the classical action $W_{-1}$ from (\ref{Clact}) by explicit summation. Additionally, we define the following auxiliary functionals for $i=1,2,3$
\begin{equation}\label{newGreen9}
g_1^i=
\frac{1}{2}\int_{\mathbb{R}^d}d^dx\int_{\mathbb{R}^d}d^dy\,J_\mu^a(x)
G^{i\,ab}_{1\,\mu\nu}(x,y)J_\nu^b(y),\,\,\,
g_0^i=
\int_{\mathbb{R}^d}d^dx\int_{\mathbb{R}^d}d^dy\,\bar{b}^{\,a}(x)
G^{i\,ab}_{0}(x,y)
b^{\,b}(y),
\end{equation}
which are actually extended versions of (\ref{oldgreen}).

\section{Two-loop contribution}
\label{sec:twol}
In this section we derive a universal formula for the two-loop contribution, which can be used for any type of regularization. For this purpose, we get an auxiliary representation, based on the modified vertices from Section \ref{sec:modver}. We want to proceed in several stages. Firstly, we write out terms for all possible combinations. Indeed, after substitution of (\ref{simp1}), (\ref{simp2}), and (\ref{newGreen5})--(\ref{newGreen7})
into the pure effective action we get three types of contributions: from the $\Gamma_3^2$-term
\begin{equation}\label{gendiag13}
-g^{2}W_{-1}[B]
\bigg(\Gamma_3^1\Gamma_3^0
\frac{J_\mu^a(0)}{6}\frac{\delta}{\delta J_\mu^a(0)}
\frac{\big(g_1^1g_1^0g_1^0\big)}{2}+
\frac{1}{2}\Gamma_3^0\Gamma_3^0
\frac{J_\mu^a(0)}{6}\frac{\delta}{\delta J_\mu^a(0)}
\frac{\big(g_1^2g_1^0g_1^0+g_1^1g_1^1g_1^0\big)}{2}
\bigg)
\bigg|_{\mathrm{1PI}\,\,\mathrm{part}}^{\substack{
\footnotesize{\mbox{UV-reg.}}\\\footnotesize{\mbox{IR-reg.}}}},
\end{equation}
from the $\Omega_3^2$-term
\begin{multline}\label{gendiag12}
-g^{2}W_{-1}[B]
\bigg(\Omega_3^1\Omega_3^0
\frac{J_\mu^a(0)}{2}\frac{\delta}{\delta J_\mu^a(0)}
\frac{\big(g_1^1g_0^0g_0^0+2g_1^0g_0^1g_1^0\big)}{2}\\+
\frac{1}{2}\Omega_3^0\Omega_3^0
\frac{J_\mu^a(0)}{2}\frac{\delta}{\delta J_\mu^a(0)}
\frac{\big(g_1^2g_0^0g_0^0+g_1^0g_0^2g_0^0+g_1^0g_0^1g_0^1\big)}{2}
\bigg)
\bigg|_{\mathrm{1PI}\,\,\mathrm{part}}^{\substack{
		\footnotesize{\mbox{UV-reg.}}\\\footnotesize{\mbox{IR-reg.}}}},
\end{multline}
and from the $\Gamma_4$-term
\begin{equation}\label{gendiag14}
g^{2}W_{-1}[B]
\bigg(\Gamma_4
\frac{J_\mu^a(0)}{4}\frac{\delta}{\delta J_\mu^a(0)}
\frac{\big(2g_1^2g_1^0+g_1^1g_1^1\big)}{2}
\bigg)
\bigg|_{\mathrm{1PI}\,\,\mathrm{part}}^{\substack{
		\footnotesize{\mbox{UV-reg.}}\\\footnotesize{\mbox{IR-reg.}}}},
\end{equation}
where we have introduced some type of ultraviolet and infrared regularizations.
All the combinations will be analyzed in the next sections. Also, let us note that in the derivation of the above formulae we have used two identities for the vertices $\Gamma_3^1$ and $\Omega_3^1$
\begin{equation}\label{commpropvert}
\big[\Gamma_3^1,J_\mu^a(0)\big]=0\,\,\,\mbox{and}\,\,\,
\big[\Omega_3^1,J_\mu^a(0)\big]=0
\,\,\,\mbox{for all $\mu$ and $a$}.
\end{equation}

\subsection{Contribution from $\Gamma_3^2$}
\label{sec:twol:gg}
\begin{figure}[h]
	\centerline{\includegraphics[width=0.74\linewidth]{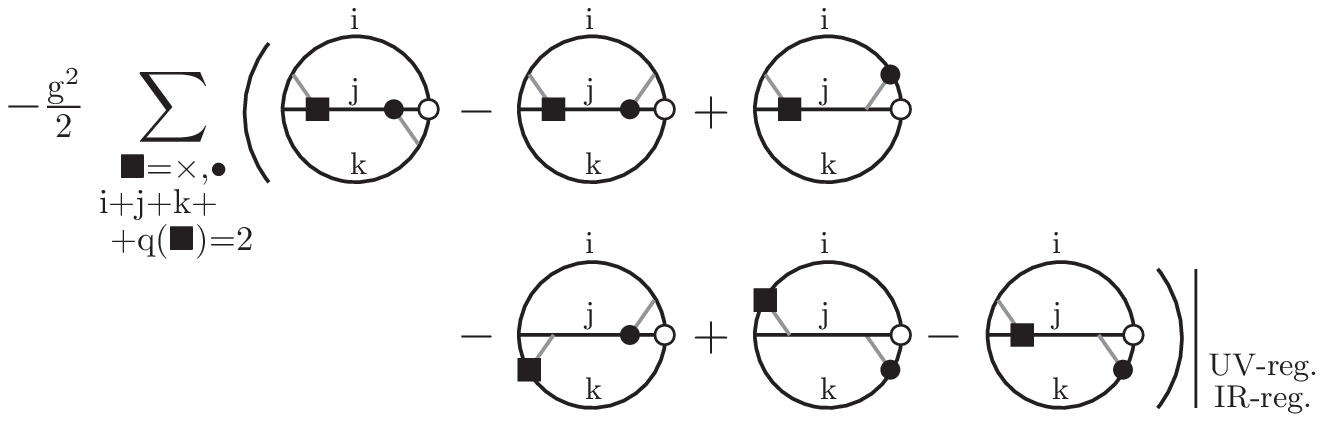}}
	\caption{Contribution from the $\Gamma_3^2$-term, where the function $\mathrm{q}$, such that $\mathrm{q}(\bullet)=0$ and $\mathrm{q}(\times)=1$, shows the degree of the background field in the corresponding vertex. The symbol $\circ$ denotes that the vertex does not contain the integration and it is considered at the zero. The numbers $\mathrm{i,j,k}$ mean the type of the propagator, see Figure \ref{propag}. }
	\label{Fig8}
\end{figure}

Let us work with formula (\ref{gendiag13}). The contributions from it can be drawn by using the Feynman diagram technique, see Figures \ref{gamma}--\ref{propag}, as it is shown in Figure \ref{Fig8}.
\begin{figure}[h]
	\centerline{\includegraphics[width=1\linewidth]{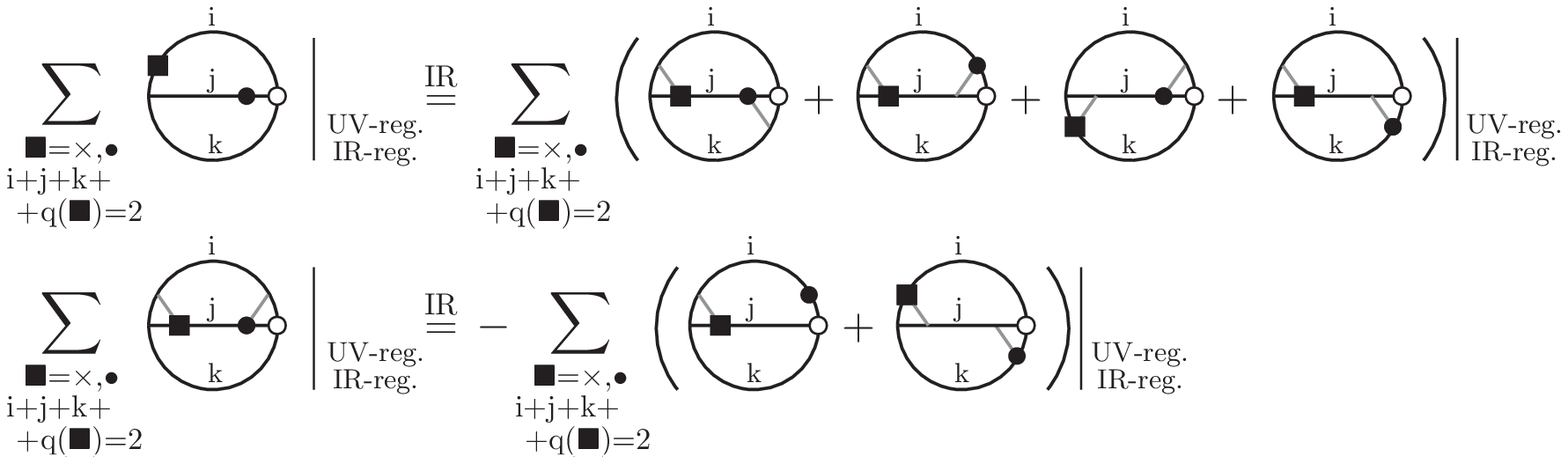}}
	\caption{Diagram equalities, where the function $\mathrm{q}$, such that $\mathrm{q}(\bullet)=0$ and $\mathrm{q}(\times)=1$, shows the degree of the background field in the corresponding vertex. The symbol $\circ$ denotes that the corresponding vertex does not contain the integration and it is considered at the zero. The numbers $\mathrm{i,j,k}$ mean the type of the propagator, see Figure \ref{propag}.}
	\label{Fig9}
\end{figure}

Thus, we have six significantly different diagrams. Fortunately, we can transform them by using two diagram relations, presented in Figure \ref{Fig9}. Such equalities were derived in the analytical form in the paper \cite{Ivanov-Kharuk-2020}, but they can be verified independently in the present restrictions.

Indeed, we need to understand, that we can transfer the element $\bullet$ or $\times$ from one line to other two with the minus sign. In other words, we should verify the rule "integration by parts". It is quite clear, because for the dot on the left hand side we can apply the usual integration by parts. For the dot on the right hand side, we also can use the integration by parts, because the integrand is a function of the difference $x-y$, and, hence, we can transfer the corresponding derivative from $y$ to $x$ and vice versa. For the crosses the property follows from equality (\ref{constprop}) for the structure constants.

\begin{figure}[h]
	\centerline{\includegraphics[width=0.56\linewidth]{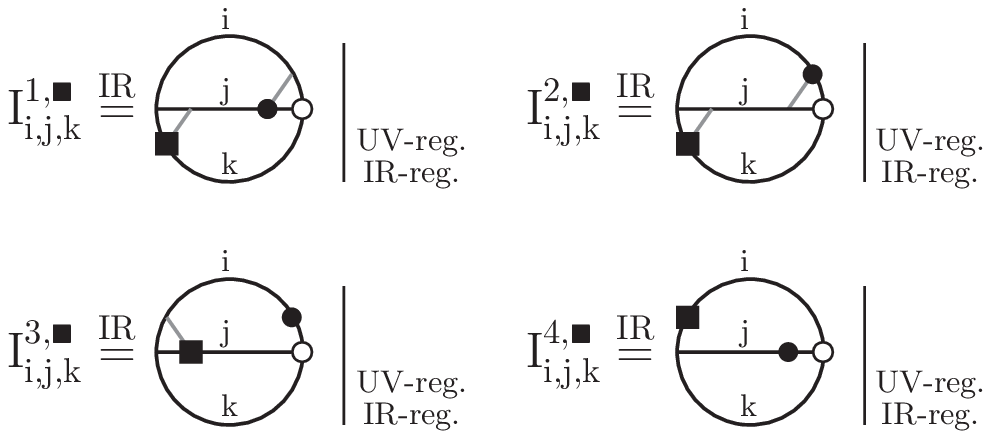}}
	\caption{The definitions of the basic graphs for the $\Gamma_3^2$-contribution, where the symbol $\scriptstyle\blacksquare$ can be replaced by $\times$ or by $\bullet$. The symbol $\circ$ denotes that the corresponding vertex does not contain the integration and it is considered at the zero. The numbers $\mathrm{i,j,k}$ mean the type of the propagator, see Figure \ref{propag}.}
	\label{Fig10}
\end{figure}

Thereby, after applying the relations from Figure \ref{Fig9} to the construction in Figure \ref{Fig8}, we can rewrite the contribution from the $\Gamma_3^2$-term in the following form
\begin{equation}\label{twoloop101}
-\sum_{n=1}^4\mathcal{J}_n,\,\,\,\mbox{where}\,\,\,
\mathcal{J}_n=\frac{\alpha_ng^2}{2}\Bigg(\sum_{\mathrm{i+j+k}=2}\mathrm{I}^{n,\bullet}_{\mathrm{i,j,k}}+\sum_{\mathrm{i+j+k}=1}\mathrm{I}^{n,\times}_{\mathrm{i,j,k}}\Bigg),
\end{equation}
and
\begin{equation}\label{twoloop102}
\alpha_1=-4,\,\,\,\alpha_2=2,\,\,\,\alpha_3=1,\,\,\,\alpha_4=1,
\end{equation}
where we have used the same notations for $\mathcal{J}_n$, as in the paper \cite{Ivanov-Kharuk-2020}, and the definitions for $\mathrm{I}^{n,\bullet}_{\mathrm{i,j,k}}$ and $\mathrm{I}^{n,\times}_{\mathrm{i,j,k}}$ are presented in Figure \ref{Fig10}.

Further, to proceed, we need to introduce some auxiliary integrals. Let $\mathrm{B}_{1/\mu}$ denotes a ball of the radius $1/\mu$, where $\mu>0$, and with the center at the origin. Then we define the following seven objects

\begin{align}\label{aped1}
\mathrm{I}_1&=\int_{\mathrm{B}_{1/\mu}}d^dx\,
f^{adc}f^{bdc}\Big(\partial_{x^\nu}G_{1\,\nu\mu}^{2\,ab}(x,y)\Big)
\Big(\partial_{y^\mu}R_0^{\phantom{2}}(x-y)\Big)
R_0^{\phantom{2}}(x-y)\bigg|_{y=0}^{\scriptsize{\mbox{IR-reg.}}},
\\\label{aped2}
\mathrm{I}_2&=\int_{\mathrm{B}_{1/\mu}}d^dx\,
f^{dca}f^{dcb}
\Big(\partial_{x^\nu}R_0^{\phantom{2}}(x-y)\Big)
\Big(\partial_{y^\mu}R_0^{\phantom{2}}(x-y)\Big)
G_{1\,\mu\nu}^{2\,ab}(x,y)\bigg|_{y=0}^{\scriptsize{\mbox{IR-reg.}}},
\\\label{aped3}
\mathrm{I}_3&=\int_{\mathrm{B}_{1/\mu}}d^dx\,
f^{dca}f^{ecb}
\Big(\partial_{x^\nu}G_{1\,\nu\mu}^{1\,de}(x,y)\Big)
\Big(\partial_{y^\sigma}R_0^{\phantom{2}}(x-y)\Big)
G_{1\,\sigma\mu}^{1\,ab}(x,y)\bigg|_{y=0}^{\scriptsize{\mbox{IR-reg.}}},\\
\label{aped4}
\mathrm{I}_4&=\int_{\mathrm{B}_{1/\mu}}d^dx\,
f^{dca}f^{ecb}
\Big(\partial_{x^\nu}G_{1\,\sigma\mu}^{1\,de}(x,y)\Big)
R_0^{\phantom{2}}(x-y)
\Big(-\partial_{x^\nu}G_{1\,\sigma\mu}^{1\,ab}(x,y)\Big)
\bigg|_{y=0}^{\scriptsize{\mbox{IR-reg.}}},
\\
\label{aped5}
\mathrm{I}_5&=\int_{\mathrm{B}_{1/\mu}}d^dx\,
f^{adc}f^{bdc}\bigg(\frac{x^\sigma}{2}
f^{age}\tilde{F}^g_{\sigma\nu}G_{1\,\nu\mu}^{1\,eb}(x,y)\bigg)
\Big(\partial_{y^\mu}R_0^{\phantom{2}}(x-y)\Big)
R_0^{\phantom{2}}(x-y)\bigg|_{y=0}^{\scriptsize{\mbox{IR-reg.}}},
\\\label{aped6}
\mathrm{I}_6&=\int_{\mathrm{B}_{1/\mu}}d^dx\,
f^{cea}f^{cbd}R_0^{\phantom{2}}(x-y)
\Big(-\partial_{x^\mu}G_{1\,\nu\mu}^{1\,eb}(x,y)\Big)
\frac{x^\sigma}{2} f^{agd}\tilde{F}^g_{\sigma\nu}R_0(x-y)
\bigg|_{y=0}^{\scriptsize{\mbox{IR-reg.}}},
\\\label{aped7}
\mathrm{I}_7&=\int_{\mathrm{B}_{1/\mu}}d^dx\,
f^{aec}f^{dbc}
\frac{x^\sigma}{2} f^{ag_1d}\tilde{F}^{g_1}_{\sigma\nu}R_0^{\phantom{2}}(x-y)
\frac{x^\rho}{2} f^{eg_2b}\tilde{F}^{g_2}_{\rho\nu}R_0^{\phantom{2}}(x-y)
R_0^{\phantom{2}}(x-y)
\bigg|_{y=0}^{\scriptsize{\mbox{IR-reg.}}},
\end{align}
where in the process of calculation we have used the explicit formulae for the Green's functions (\ref{newGreen5})--(\ref{newGreen7}).

Then, our main idea is to express the diagrams from Figure \ref{Fig10} in terms of the last integrals. It is a quite simple and boring computations, so we present only the final compliance table
\begin{center}
\begin{tabular}{l l l l}
$\mathrm{I_{2,0,0}^{1,\bullet}\stackrel{IR}{=}-I_2^{\phantom{0}}},$
&$\mathrm{I_{2,0,0}^{2,\bullet}\stackrel{IR}{=}-}d\mathrm{I_1^{\phantom{0}}},$
&$\mathrm{I_{2,0,0}^{3,\bullet}\stackrel{IR}{=}-}d\mathrm{I_1^{\phantom{0}}-}d\mathrm{I_5^{\phantom{0}}},$
&$\mathrm{I_{2,0,0}^{4,\bullet}\stackrel{IR}{=}-I_1^{\phantom{0}}},$ \\
$\mathrm{I_{0,2,0}^{1,\bullet}\stackrel{IR}{=}-I_1^{\phantom{0}}-I_5^{\phantom{0}}},$
&$\mathrm{I_{0,2,0}^{2,\bullet}\stackrel{IR}{=}-}d\mathrm{I_2^{\phantom{0}}},$
&$\mathrm{I_{0,2,0}^{3,\bullet}\stackrel{IR}{=}-}d\mathrm{I_1^{\phantom{0}}},$
&$\mathrm{I_{0,2,0}^{4,\bullet}\stackrel{IR}{=}-I_1^{\phantom{0}}-I_5^{\phantom{0}}},$ \\
$\mathrm{I_{0,0,2}^{1,\bullet}\stackrel{IR}{=}-I_1^{\phantom{0}}},$
&$\mathrm{I_{0,0,2}^{2,\bullet}\stackrel{IR}{=}-}d\mathrm{I_1^{\phantom{0}}},$
&$\mathrm{I_{0,0,2}^{3,\bullet}\stackrel{IR}{=}-}d\mathrm{I_2^{\phantom{0}}},$
&$\mathrm{I_{0,0,2}^{4,\bullet}\stackrel{IR}{=}-I_2^{\phantom{0}}},$ \\
$\mathrm{I_{1,1,0}^{1,\bullet}\stackrel{IR}{=}I_3^{\phantom{0}}+\frac{1}{2}I_5^{\phantom{0}}},$
&$\mathrm{I_{1,1,0}^{2,\bullet}\stackrel{IR}{=}}\,\,\frac{d}{2}\mathrm{I_5^{\phantom{0}}},$
&$\mathrm{I_{1,1,0}^{3,\bullet}\stackrel{IR}{=}0},$
&$\mathrm{I_{1,1,0}^{4,\bullet}\stackrel{IR}{=}}\,\,\frac{1}{d}\mathrm{I_4^{\phantom{0}}-I_6^{\phantom{0}}},$\\
$\mathrm{I_{1,0,1}^{1,\bullet}\stackrel{IR}{=}-I_3^{\phantom{0}}},$
&$\mathrm{I_{1,0,1}^{2,\bullet}\stackrel{IR}{=}-I_4^{\phantom{0}}},$
&$\mathrm{I_{1,0,1}^{3,\bullet}\stackrel{IR}{=}0},$
&$\mathrm{I_{1,0,1}^{4,\bullet}\stackrel{IR}{=}-I_3^{\phantom{0}}},$ \\
$\mathrm{I_{0,1,1}^{1,\bullet}\stackrel{IR}{=}-}\frac{1}{d}\mathrm{I_4^{\phantom{0}}+I_6^{\phantom{0}}},$
&$\mathrm{I_{0,1,1}^{2,\bullet}\stackrel{IR}{=}0},$
&$\mathrm{I_{0,1,1}^{3,\bullet}\stackrel{IR}{=}-}d\mathrm{I_3^{\phantom{0}}},$
&$\mathrm{I_{0,1,1}^{4,\bullet}\stackrel{IR}{=}-I_3^{\phantom{0}}-\frac{1}{2}I_5^{\phantom{0}}},$ \\
$\mathrm{I_{1,0,0}^{1,\times}\stackrel{IR}{=}\frac{1}{2}I_5^{\phantom{0}}},$
&$\mathrm{I_{1,0,0}^{2,\times}\stackrel{IR}{=}-}d\mathrm{I_7^{\phantom{0}}},$
&$\mathrm{I_{1,0,0}^{3,\times}\stackrel{IR}{=}-}d\mathrm{I_6^{\phantom{0}}-}d\mathrm{I_7^{\phantom{0}}},$
&$\mathrm{I_{1,0,0}^{4,\times}\stackrel{IR}{=}-I_5^{\phantom{0}}},$ \\
$\mathrm{I_{0,1,0}^{1,\times}\stackrel{IR}{=}-I_6^{\phantom{0}}-I_7^{\phantom{0}}},$
&$\mathrm{I_{0,1,0}^{2,\times}\stackrel{IR}{=}}\,\,\frac{d}{2}\mathrm{I_5^{\phantom{0}}},$
&$\mathrm{I_{0,1,0}^{3,\times}\stackrel{IR}{=}0},$
&$\mathrm{I_{0,1,0}^{4,\times}\stackrel{IR}{=}-I_6^{\phantom{0}}-I_7^{\phantom{0}}},$ \\
$\mathrm{I_{0,0,1}^{1,\times}\stackrel{IR}{=}I_5^{\phantom{0}}},$
&$\mathrm{I_{0,0,1}^{2,\times}\stackrel{IR}{=}0},$
&$\mathrm{I_{0,0,1}^{3,\times}\stackrel{IR}{=}0},$
&$\mathrm{I_{0,0,1}^{4,\times}\stackrel{IR}{=}-\frac{1}{2}I_5^{\phantom{0}}}.$
\end{tabular}
\end{center}

Using the last table and formula (\ref{twoloop101}), we obtain immediately the following result
\begin{align}\label{twoloop103}
\mathcal{J}_1\stackrel{\mathrm{IR}}{=}&g^2\bigg(4\mathrm{I}_1^{\phantom{0}}+2\mathrm{I}_2^{\phantom{0}}+\frac{2}{d}\mathrm{I}_4^{\phantom{0}}-2\mathrm{I}_5^{\phantom{0}}+2\mathrm{I}_7^{\phantom{0}}
\bigg),
\\\label{twoloop104}
\mathcal{J}_2\stackrel{\mathrm{IR}}{=}&g^2\Big(-2d\mathrm{I}_1^{\phantom{0}}-d\mathrm{I}_2^{\phantom{0}}-\mathrm{I}_4^{\phantom{0}}+d\mathrm{I}_5^{\phantom{0}}-d\mathrm{I}_7^{\phantom{0}}
\Big),
\\\label{twoloop105}
\mathcal{J}_3\stackrel{\mathrm{IR}}{=}&g^2\bigg(-d\mathrm{I}_1^{\phantom{0}}-\frac{d}{2}\mathrm{I}_2^{\phantom{0}}-\frac{d}{2}\mathrm{I}_3^{\phantom{0}}-\frac{d}{2}\mathrm{I}_5^{\phantom{0}}-\frac{d}{2}\mathrm{I}_6^{\phantom{0}}-\frac{d}{2}\mathrm{I}_7^{\phantom{0}}
\bigg),
\\\label{twoloop106}
\mathcal{J}_4\stackrel{\mathrm{IR}}{=}&g^2\bigg(-\mathrm{I}_1^{\phantom{0}}-\frac{1}{2}\mathrm{I}_2^{\phantom{0}}-\mathrm{I}_3^{\phantom{0}}+\frac{1}{2d}\mathrm{I}_4^{\phantom{0}}-\frac{3}{2}\mathrm{I}_5^{\phantom{0}}-
\mathrm{I}_6^{\phantom{0}}-\frac{1}{2}\mathrm{I}_7^{\phantom{0}}
\bigg),
\\\label{twoloop107}
-\sum_{n=1}^4\mathcal{J}_n\stackrel{\mathrm{IR}}{=}&
g^2\bigg((3d-3)\mathrm{I}_1^{\phantom{0}}+\frac{(3d-3)}{2}\mathrm{I}_2^{\phantom{0}}+\frac{(d+2)}{2}\mathrm{I}_3^{\phantom{0}}\\\nonumber&\,\,\,\,\,\,\,\,\,\,
\,\,\,\,\,\,\,\,\,\,\,\,\,\,\,
+\frac{(2d-5)}{2d}\mathrm{I}_4^{\phantom{0}}
+\frac{(7-d)}{2}\mathrm{I}_5^{\phantom{0}}+\frac{(d+2)}{2}
\mathrm{I}_6^{\phantom{0}}+\frac{(3d-3)}{2}\mathrm{I}_7^{\phantom{0}}
\bigg).
\end{align}

Let us note one more time that in all calculations, we kept the parameter of the dimension to have the ability to study the case of dimensional regularization. In all other situations (without deformation of the dimension of the space) we can substitute $d=4$.

\subsection{Contribution from $\Omega_3^2$}
\label{sec:twol:oo}
Now we are going to find the divergence in the $\Omega_3^2$-term using formula (\ref{gendiag12}). Actually, we need to repeat all steps, that have been undertaken in the case of the $\Gamma_3^2$-term, but in simplified form, because we have only one type of the diagram.

\begin{figure}[h]
	\centerline{\includegraphics[width=0.6\linewidth]{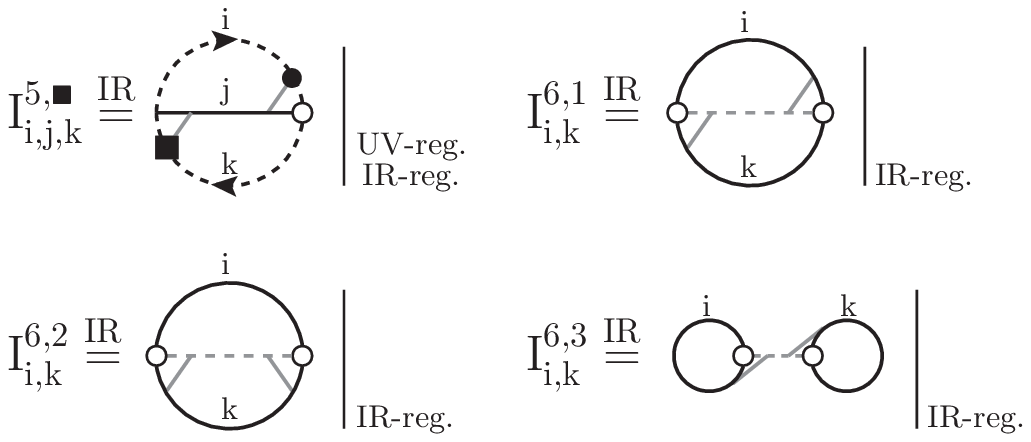}}
	\caption{Contributions from $\Omega_3^2$ and $\Gamma_4^{\phantom{2}}$, where the function $\mathrm{q}$, such that $\mathrm{q}(\bullet)=0$ and $\mathrm{q}(\times)=1$, shows the degree of the background field in the corresponding vertex. The symbol $\circ$ denotes that the corresponding vertex does not contain the integration and it is considered at the zero. The numbers $\mathrm{i,j,k}$ mean the type of the propagator, see Figure \ref{propag}.}
	\label{Fig11}
\end{figure}

Indeed, in this case the corresponding contribution can be rewritten in the form
\begin{equation}\label{twoloop108}
-\mathcal{J}_5\stackrel{\mathrm{IR}}{=}\frac{g^2}{2}\Bigg(\sum_{\mathrm{i+j+k}=2}\mathrm{I}^{5,\bullet}_{\mathrm{i,j,k}}+\sum_{\mathrm{i+j+k}=1}\mathrm{I}^{5,\times}_{\mathrm{i,j,k}}\Bigg),
\end{equation}
where the objects $\mathrm{I}^{5,\bullet}_{\mathrm{i,j,k}}$ and $\mathrm{I}^{5,\times}_{\mathrm{i,j,k}}$ are depicted in Figure \ref{Fig11}, and the form factor $\mathcal{J}_5$ is selected in the same form, as it was in the work \cite{Ivanov-Kharuk-2020}.

To proceed we need to introduce one more type of integral in addition to the ones from (\ref{aped1})--(\ref{aped7}) in the form
\begin{equation}\label{aped8}
\mathrm{I}_8=\int_{\mathrm{B}_{1/\mu}}d^dx\,
f^{dca}f^{dcb}
\Big(\partial_{y^\mu}R_0^{\phantom{2}}(x-y)\Big)
R_0^{\phantom{2}}(x-y)
\Big(\partial_{x^\mu}G_{0}^{2\,ab}(x,y)\Big)
\bigg|_{y=0}^{\scriptsize{\mbox{IR-reg.}}}.
\end{equation}

Then we give the corresponding table with relations, which has the form
\begin{center}
	\begin{tabular}{l l l}
		$\mathrm{I_{2,0,0}^{5,\bullet}\stackrel{IR}{=}-I_8^{\phantom{0}}},$
		&$\mathrm{I_{1,1,0}^{5,\bullet}\stackrel{IR}{=}\frac{1}{2}I_5^{\phantom{0}}},$
		&$\mathrm{I_{1,0,0}^{5,\times}\stackrel{IR}{=}-I_7^{\phantom{0}}},$ \\
		$\mathrm{I_{0,2,0}^{5,\bullet}\stackrel{IR}{=}-I_2^{\phantom{0}}},$
		&$\mathrm{I_{1,0,1}^{5,\bullet}\stackrel{IR}{=}0},$
		&$\mathrm{I_{0,1,0}^{5,\times}\stackrel{IR}{=}\frac{1}{2}I_5^{\phantom{0}}},$ \\
		$\mathrm{I_{0,0,2}^{5,\bullet}\stackrel{IR}{=}-I_8^{\phantom{0}}},$
		&$\mathrm{I_{0,1,1}^{5,\bullet}\stackrel{IR}{=}0},$
		&$\mathrm{I_{0,0,1}^{5,\times}\stackrel{IR}{=}0}.$
	\end{tabular}
\end{center}

Hence, after summing all terms we get the answer depending only on four types of the integrals
\begin{equation}\label{twoloop109}
-\mathcal{J}_5\stackrel{\mathrm{IR}}{=}g^2\Bigg(
-\frac{1}{2}\mathrm{I}_2^{\phantom{0}}+\frac{1}{2}\mathrm{I}_5^{\phantom{0}}-
\frac{1}{2}\mathrm{I}_7^{\phantom{0}}-\mathrm{I}_8^{\phantom{0}}
\Bigg).
\end{equation}

\subsection{Contribution from $\Gamma_4$}
\label{sec:twol:g}
The last divergence follows from the $\Gamma_4$-term, see formula (\ref{gendiag14}). In the Feynman diagram language, it can be formulated by using the element in Figure \ref{gamma4}. Hence, the contribution can be decomposed on the basis of three diagrams, depicted in Figure \ref{Fig11}, and has the following view
\begin{equation}\label{twoloop110}
-\mathcal{J}_6\stackrel{\mathrm{IR}}{=}\frac{g^2}{4}
\sum_{\mathrm{i+k}=2}\Big(
\mathrm{I}^{6,1}_{\mathrm{i,k}}-\mathrm{I}^{6,2}_{\mathrm{i,k}}+\mathrm{I}^{6,3}_{\mathrm{i,k}}
\Big),
\end{equation}
where we again used the notation convenient for comparison with the work \cite{Ivanov-Kharuk-2020}.

Further, introducing two auxiliary constructions
\begin{equation}\label{aped9}
\mathrm{I}_9=c_2^2R_1^2(x)
\Big|_{x=0}^{\scriptsize{\mbox{IR-reg.}}},
\,\,\,
\mathrm{I}_{10}=c_2^2R_0^{\phantom{2}}(x)R_2^{\phantom{2}}(x)
	\Big|_{x=0}^{\scriptsize{\mbox{IR-reg.}}},
\end{equation}
we can write out the table
\begin{center}
	\begin{tabular}{l l l}
		$\mathrm{I_{2,0}^{6,1}\stackrel{IR}{=}}\,-\frac{5}{3}\mathrm{I_{10}^{\phantom{0}}},$
		&$\mathrm{I_{2,0}^{6,2}\stackrel{IR}{=}}\,-\frac{20}{3}\mathrm{I_{10}^{\phantom{0}}},$
		&$\mathrm{I_{2,0}^{6,3}\stackrel{IR}{=}0},$ \\
		$\mathrm{I_{1,1}^{6,1}\stackrel{IR}{=}}\,\,2\mathrm{I_9^{\phantom{0}}},$
		&$\mathrm{I_{1,1}^{6,2}\stackrel{IR}{=}\,0},$
		&$\mathrm{I_{1,1}^{6,3}\stackrel{IR}{=}}\,4\mathrm{I_9^{\phantom{0}}},$ \\
		$\mathrm{I_{0,2}^{6,1}\stackrel{IR}{=}}\,-\frac{5}{3}\mathrm{I_{10}^{\phantom{0}}},$
		&$\mathrm{I_{0,2}^{6,2}\stackrel{IR}{=}}\,-\frac{20}{3}\mathrm{I_{10}^{\phantom{0}}},$
		&$\mathrm{I_{0,2}^{6,3}\stackrel{IR}{=}0},$
	\end{tabular}
\end{center}
and the answer in the form
\begin{equation}\label{twoloop111}
	-\mathcal{J}_6\stackrel{\mathrm{IR}}{=}g^2\Bigg(
	\frac{3}{2}\mathrm{I}_9^{\phantom{0}}+\frac{5}{2}\mathrm{I}_{10}^{\phantom{0}}
	\Bigg).
\end{equation}
\subsection{Quantum equation of motion}
\label{sec:twol:quant}
In this section we want to discuss briefly a quantum equation of motion. This leads to a counterterm, that appears in an effective action after the renormalization of the pure effective action. Such way gives the ability to compare answers in the case of the dimensional regularization with the results obtained earlier. 

First of all, let us derive it in the first powers of the coupling constant. As it was noted in the works \cite{12,23,Ivanov-Russkikh}, we need to consider the diagram "glasses"
\begin{equation}
\label{qemm1}
\bigg(-\frac{(1-g\theta)^2}{2g^2}\Gamma_1^2-(1-g\theta)
\Gamma_1^{\phantom{2}}\Gamma_3^{\phantom{2}}+(1-g\theta)
\Gamma_1^{\phantom{2}}\Omega_3^{\phantom{2}}+\mathcal{O}\big(g^2\big)\bigg)
Z[J,b,\bar{b}\,]\Big|_{J_\mu=b=\bar{b}=0}=0,
\end{equation}
where we have used the notations from Section \ref{sec:bas:ym}, see formulae (\ref{versh1}), (\ref{versh2}), and (\ref{oldgreen}). Also, $\theta=\theta(g)$ is the second renormalization constant for the Yang--Mills theory, that will be discussed below.

Further, we can proceed in two different ways: find a quadratic form, as it was made in \cite{12}, from which the equation follows, or find a variation by the vertex $\Gamma_1$. Both ways are possible and give the same equality
\begin{equation}
\label{qemm2}
-\frac{(1-g\theta)}{g}\Gamma_1-g\Gamma_3g_1+g\Omega_3g_0+\mathcal{O}\big(g^2\big)=0.
\end{equation}

Left hand side of the last relation is the functional of the auxiliary arbitrary smooth field $J_\mu(x)$. It means that we can consider only the integrand. Hence, using the integration by parts to remove the derivative from the field $J_\mu(x)$, we obtain
\begin{align}
\label{qemm3}
-\frac{(1-g\theta)}{g}\big(-D_{x_\mu}^{cb}F^b_{\mu\nu}(x)\big)
&+gf^{abc}D_{x_\mu}^{ae}G_{1\nu\nu}^{\,\,\,eb}(x,y)\big|_{y=x}
+gf^{abc}D_{x_\nu}^{ae}G_{1\nu\mu}^{\,\,\,eb}(x,y)\big|_{y=x}\\&\nonumber
-2gf^{abc}D_{x_\nu}^{ae}G_{1\mu\nu}^{\,\,\,eb}(x,y)\big|_{y=x}
-gf^{abc}D_{x_\mu}^{ae}G_{0}^{eb}(x,y)\big|_{y=x}
+\mathcal{O}\big(g^2\big)=0,
\end{align}
where the second, the third, and the fourth terms follow from $-g\Gamma_3g_1$.

We are interested only in the part proportional to the classical equation of motion $D_{x_\mu}^{cb}F^b_{\mu\nu}(x)$. It is quite easy to see that for calculations we can use only the second term from (\ref{SD9}), where the first Seeley--DeWitt coefficients have the following form, see \cite{33,Ivanov-Kharuk-2020},
\begin{align}
\label{qemm4}
a_{1\mu\nu}(x,y)&=2F_{\mu\nu}(y)+
(x-y)^\sigma\bigg(
\nabla_{y_\sigma}F_{\mu\nu}(y)+
\frac{\delta_{\mu\nu}}{6}
\nabla_{y_\rho}F_{\sigma\rho}(y)-2
B_{\sigma}F_{\mu\nu}(y)
\bigg)+\mathcal{O}\big(|x-y|^2\big),\\\label{qemm5}
a_{0}(x,y)&=
\frac{1}{6}(x-y)^\sigma
\nabla_{y_\rho}F_{\sigma\rho}(y)
+\mathcal{O}\big(|x-y|^2\big),
\end{align}
where $\nabla_{y_\sigma}\cdot=[D_{y_\sigma},\,\cdot\,\,]$.
Then, we can write out one more auxiliary formula
\begin{equation}
\label{qemm6}
f^{abc}\nabla_{y_\mu}^{ae}f^{edb}F_{\nu\rho}^d(y)=-c_2
D_{y_\mu}^{ca}F_{\nu\rho}^a(y),
\end{equation}
where we have used formula (\ref{constprop}). Therefore, after applying the covariant derivative to (\ref{qemm4})--(\ref{qemm5}) and substituting relation (\ref{qemm6}), we get
\begin{align}
\label{qemm7}
f^{abc}D_{x_\mu}^{ae}a_{1\nu\rho}^{\,\,\,eb}(x,y)\big|_{y=x}&=-c_2
\bigg(D_{x_\mu}^{ca}F_{\nu\rho}^a(x)+\frac{\delta_{\nu\rho}}{6}
D_{x_\sigma}^{ca}F_{\mu\sigma}^a(x)\bigg),\\\label{qemm8}
f^{abc}D_{x_\mu}^{ae}a_{1}^{eb}(x,y)\big|_{y=x}&=-
\frac{c_2}{6}
D_{x_\sigma}^{ca}F_{\mu\sigma}^a(x).
\end{align}

Hence, equation (\ref{qemm3}) can be rewritten in the form
\begin{equation}
\label{qemm9}
D_{x_\mu}^{cb}F^b_{\mu\nu}(x)
\bigg(\frac{1}{g}-\theta-\frac{8}{3}c_2gR_1(z)
\Big|^{\scriptsize{\mbox{IR-reg.}}}_{z=0}\bigg)+\ldots=0,
\end{equation}
where the dots denote the terms, we are not interested in. They are either without the logarithmic singularity or with higher degrees of the coupling constant.

Now we need to use the general renormalization theory  for the Yang--Mills theory, see the works \cite{3,12}. To find a form factor for a counterterm in the two-loop calculations, we need to make one-loop renormalizations of the effective action and the quantum equation of motion. Let us do it in stages.

As it was noted in the papers \cite{16,17}, the first loop contains a divergent part, that can be represented in the following form after some type of infrared regularization
\begin{equation}
\label{qemm10}
-\frac{11}{3}\frac{c_2}{4}W_{-1}R_1(z)
\Big|^{\scriptsize{\mbox{IR-reg.}}}_{z=0}.
\end{equation}

Hence, to avoid the presence of the divergence, according to the general theory we need to change the coupling constant as
\begin{equation}
\label{qemm11}
\frac{1}{g^2_{\phantom{\scriptsize{\mbox{1}}}}}\to\frac{1}{g^2_{\phantom{\scriptsize{\mbox{1}}}}}
+\frac{11}{3}c_2R_1(z)
\Big|^{\scriptsize{\mbox{IR-reg.}}}_{z=0}.
\end{equation}

After that we can move on to the quantum equation in the form (\ref{qemm9}). Firstly, we replace the coupling constant $g$ by the renormalized one. It means that the expression in parentheses from (\ref{qemm9}) has the view
\begin{equation}
\label{qemm12}
\frac{1}{g}-\theta-\frac{5}{6}c_2gR_1(z)
\Big|^{\scriptsize{\mbox{IR-reg.}}}_{z=0}.
\end{equation}

Then, according to the main idea of the renormalization procedure, we make one more shift

\begin{equation}
\label{qemm13}
\theta\to\theta
-\frac{5}{6}c_2gR_1(z)
\Big|^{\scriptsize{\mbox{IR-reg.}}}_{z=0}
\end{equation}
to cancel the divergence. This transformation leads to the appearance of an additional vertex with two external lines
\begin{equation}
\label{qemm15}
\Gamma_2=-R_1(z)
\Big|^{\scriptsize{\mbox{IR-reg.}}}_{z=0}\,
\frac{5g^2c_2}{6}
\int_{\mathbb{R}^d}d^dx\,
\frac{\delta}{\delta J_\nu^{\,a}(x)}
\overrightarrow{D}^{ab}_{x_{\nu}}\overrightarrow{D}^{bc}_{x_{\mu}}
\frac{\delta}{\delta J_\mu^{\,c}(x)},
\end{equation}
which does not appear in the pure effective action and which should be included in the exponential from (\ref{highloops}). Then, the pure effective action after the one-loop renormalization get
the following counterterm to the two-loop contribution
\begin{equation}
\label{qemm14}
-\mathcal{J}_7\stackrel{\mathrm{IR}}{=}
R_1(z)
\Big|^{\scriptsize{\mbox{IR-reg.}}}_{z=0}\,
\frac{5g^2c_2}{6}
\int_{\mathbb{R}^d}d^dx\,\overrightarrow{D}^{ab}_{x_{\nu}}\overrightarrow{D}^{bc}_{x_{\mu}}
G_{1\mu\nu}^{\,\,\,ca}(x,y)\Big|^{\scriptsize{\mbox{IR-reg.}}}_{y=x}.
\end{equation}

\section{Some types of regularization}
\label{sec:exam:}
\subsection{Dimensional regularization}
\label{sec:exam:dr}
Now we are going to apply formula (\ref{twoloop112}) in the case of dimensional regularization. As it was noted above, we preserved the parameter of dimension, see formulae in Section \ref{sec:answtwoloop}, hence, it is possible. Of course, we are not going to explain all the subtleties of the regularization, but we give only required information for our computations. Detailed information about the introduction of the regularization can be found in the papers \cite{5,19,12}.

First of all we should draw the attention that the dimension of the space is not an integer. It is equal to $d=4-\varepsilon$, where $\varepsilon$ is a dimensionless parameter of the regularization. It means that we can obtain the standard theory in the following limit $\varepsilon\to+0$.

Then, according to formulae from Section \ref{sec:answtwoloop}, we need to introduce the regularized versions of the $R_i(x)$-functions, where $i=0,1,2$. They have the following definitions, see the first part in Figure \ref{regs},
\begin{equation}\label{dmreg1}
R_0^{\varepsilon}(x)=\frac{\Gamma(d/2-1)}{4\pi^{d/2}}|x|^{2-d},\,\,\,
R_1^{\varepsilon}(x)=\frac{1}{16\pi^2}\bigg(\frac{2\mu^{-\varepsilon}}{\varepsilon}+
\frac{\Gamma(d/2-2)}{\pi^{d/2-2}}|x|^{4-d}\bigg),
\end{equation}
\begin{equation}\label{dmreg2}
R_2^{\varepsilon}(x)=\frac{1}{32\pi^2}\bigg(-\frac{|x|^2\mu^{-\varepsilon}}{\varepsilon}+
\frac{\Gamma(d/2-3)}{2\pi^{d/2-2}}|x|^{6-d}\bigg),
\end{equation}
where $\mu$ is an auxiliary parameter to keep the dimension of the constructions. It has a finite value.

It is quite easy to verify that after removing the regularization $\varepsilon\to+0$, we obtain the standard functions from (\ref{SD10}) with additional terms
\begin{equation}\label{dmreg7}
R_0^{\varepsilon}(x)\to R_0(x),\,\,\,
R_1^{\varepsilon}(x)\to R_1(x)-\frac{\gamma+\ln(\pi)}{(4\pi)^2},\,\,\,
R_2^{\varepsilon}(x)\to R_2(x)+\frac{\gamma+\ln(\pi)}{4(4\pi)^2}|x|^2.
\end{equation}
The last additional terms can not be considered, because they are from the $\mathcal{ZM}$-term, and therefore, according to the results of the paper \cite{Kharuk-2021}, they are not affecting the divergent part of the two-loop contribution.

Then, for the simplicity of calculations, we present some useful properties of the last regularized functions
\begin{equation}\label{dmreg3}
-\partial_{x_\mu}\partial_{x^\mu}R_0^{\varepsilon}(x)=\delta^d(x),\,\,\,
-\partial_{x_\mu}\partial_{x^\mu}R_1^{\varepsilon}(x)=R_0^{\varepsilon}(x),\,\,\,
-\partial_{x_\mu}\partial_{x^\mu}R_2^{\varepsilon}(x)=2R_1^{\varepsilon}(x)-\frac{\mu^{-\varepsilon}}{16\pi^2},
\end{equation}
\begin{equation}\label{dmreg4}
-2\partial_{x_\mu}R_1^{\varepsilon}(x)=x^\mu R_0^{\varepsilon}(x),\,\,\,
-2\partial_{x_\mu}R_2^{\varepsilon}(x)=x^\mu R_1^{\varepsilon}(x),
\end{equation}
\begin{equation}\label{dmreg5}
x^\mu\partial_{x^\mu}R_0^{\varepsilon}(x)=(2-d)R_0^{\varepsilon}(x),\,\,\,
|x|^2\partial_{x^\mu}R_0^{\varepsilon}(x)\partial_{x_\mu}R_0^{\varepsilon}(x)=(2-d)^2R_0^{\varepsilon}(x)R_0^{\varepsilon}(x).
\end{equation}

By using the last properties and definitions (\ref{dmreg1}) and (\ref{dmreg2}), we can simplify the integrals (\ref{aped17})--(\ref{aped16}) and find some relations among them. They have the form
\begin{equation}\label{dmreg8}
\mathrm{I}_1\stackrel{\mathrm{IR}}{=}\bigg(\frac{1}{6}-\frac{(d-24)}{24}\bigg)\mathrm{I}_3+\frac{(2-d)}{12d}\mathrm{I}_4+\frac{(24-d)}{d}\mathrm{I}_{\mathrm{aux}},\,\,\,
\mathrm{I}_2\stackrel{\mathrm{IR}}{=}-\mathrm{I}_1,
\end{equation}
\begin{equation}\label{dmreg6}
\mathrm{I}_5\stackrel{\mathrm{IR}}{=}-\mathrm{I}_3,\,\,\,
\mathrm{I}_6\stackrel{\mathrm{IR}}{=}\frac{1}{2d}\mathrm{I}_4,\,\,\,
\mathrm{I}_7\stackrel{\mathrm{IR}}{=}-\frac{1}{4d}\mathrm{I}_4,\,\,\,
\mathrm{I}_8\stackrel{\mathrm{IR}}{=}\bigg(\frac{1}{6}-\frac{d}{24}\bigg)\mathrm{I}_3+\frac{(2-d)}{12d}\mathrm{I}_4-\mathrm{I}_{\mathrm{aux}},
\end{equation}
where actually we need to calculate only two integrals
\begin{equation}\label{dmreg9}
\mathrm{I}_3=\frac{(2-d)c_2^2}{d}\int
d^dx\,
R_0^{\varepsilon}(x)R_0^{\varepsilon}(x)R_1^{\varepsilon}(x),\,\,\,
\mathrm{I}_4=-2c_2^2\int
d^dx\,
R_0^{\varepsilon}(x)\Big(\partial_{x^\mu}R_1^{\varepsilon}(x)\Big)
\Big(\partial_{x^\mu}R_1^{\varepsilon}(x)\Big),
\end{equation}
and one auxiliary integral
\begin{equation}\label{dmreg10}
\mathrm{I}_{\mathrm{aux}}=\frac{(2-d)c_2^2}{2^83\pi^2}\int
d^dx\,
R_0^{\varepsilon}(x)R_0^{\varepsilon}(x).
\end{equation}

From the last manipulations we see that indeed we need to use only three basic relations. They have the form, see \cite{12},
\begin{equation}\label{dmreg11}
R_0^{\varepsilon}(x)R_0^{\varepsilon}(x)R_1^{\varepsilon}(x)
\sim\frac{\mu^{-2\varepsilon}}{(4\pi)^4}\bigg(\frac{2}{\varepsilon^2}+\frac{1}{\varepsilon}\bigg)\delta^d(x)
,
\end{equation}
\begin{equation}\label{dmreg12}
R_0^{\varepsilon}(x)\Big(\partial_{x^\mu}R_1^{\varepsilon}(x)\Big)
\Big(\partial_{x^\mu}R_1^{\varepsilon}(x)\Big)
\sim\frac{1}{(4\pi)^4}\frac{d}{4\varepsilon}\delta^d(x)
,\,\,\,
R_0^{\varepsilon}(x)R_0^{\varepsilon}(x)
\sim\frac{1}{8\pi^2}\frac{1}{\varepsilon}\delta^d(x).
\end{equation}

Hence, after the preparations we can easily write out the integrals $\mathrm{I}_1$--$\mathrm{I}_{10}$ and find the two-loop contribution. All answers can be found in the result tables in Section \ref{sec:explicit}.

\subsection{Cutoff regularization}
\label{sec:exam:cut}

\begin{figure}[h]
\includegraphics[width=0.33\linewidth]{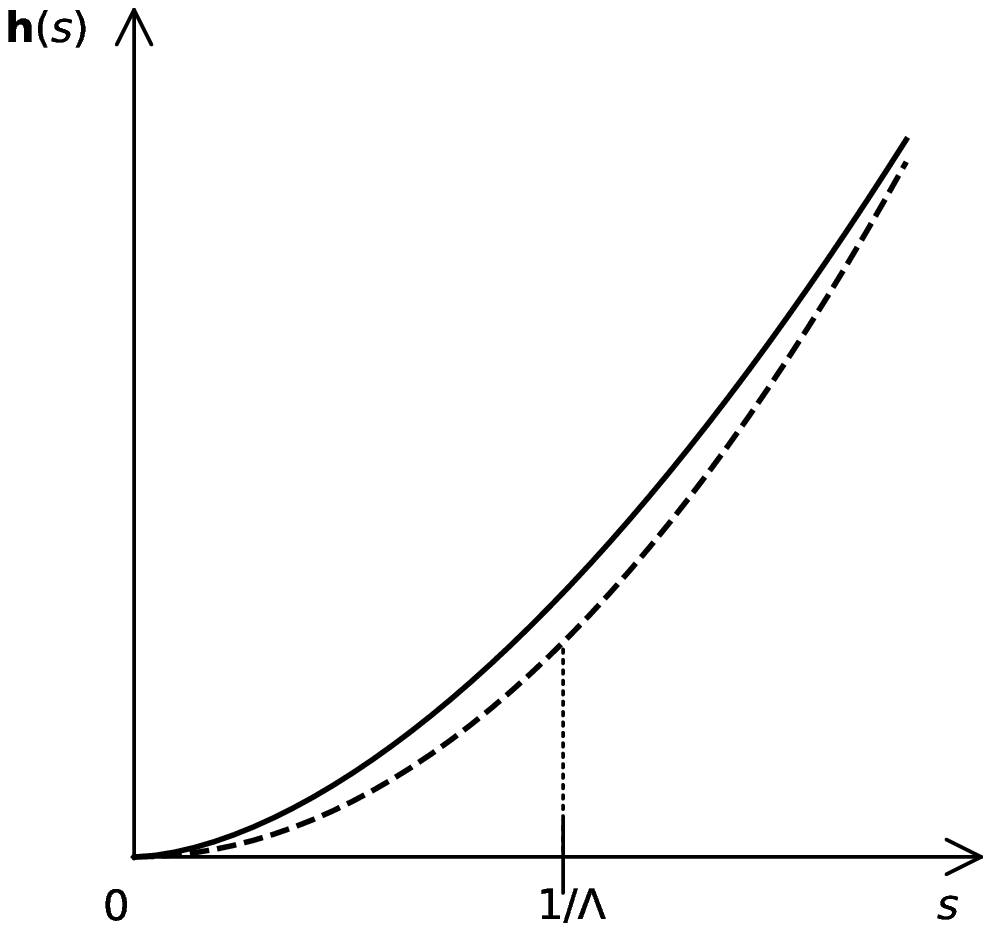}
\includegraphics[width=0.33\linewidth]{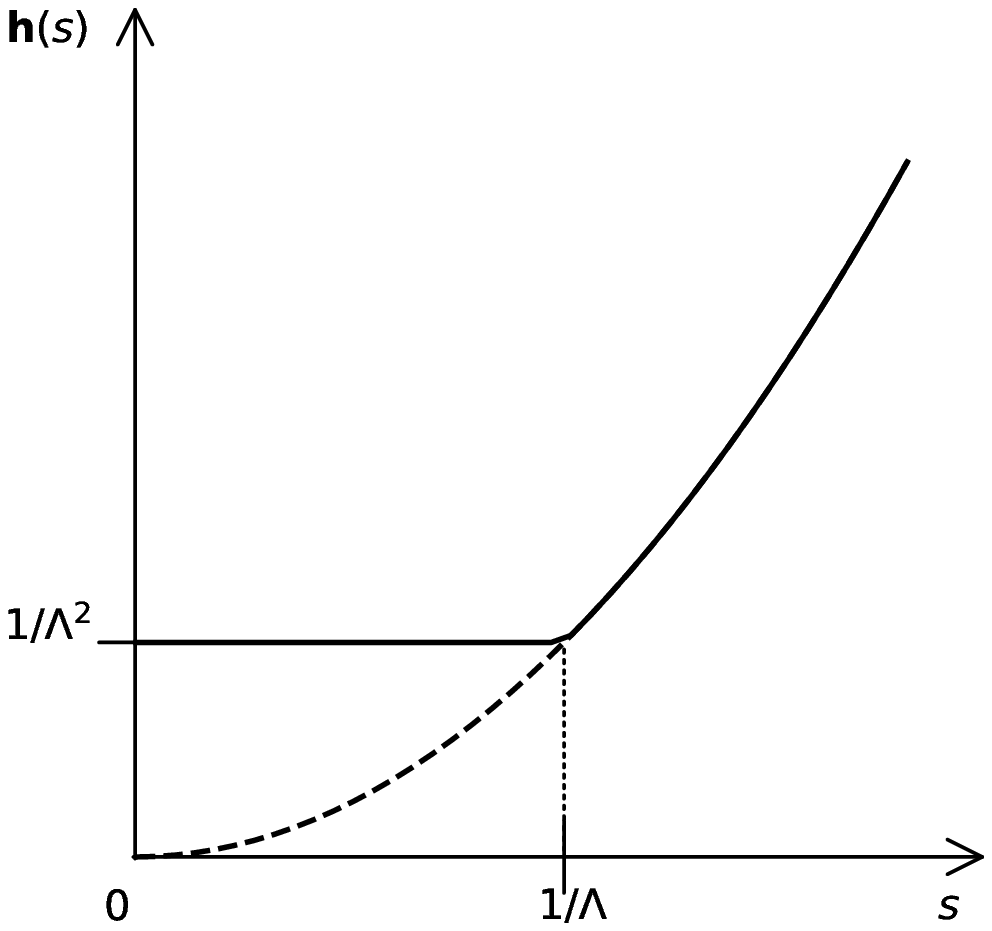}
\includegraphics[width=0.33\linewidth]{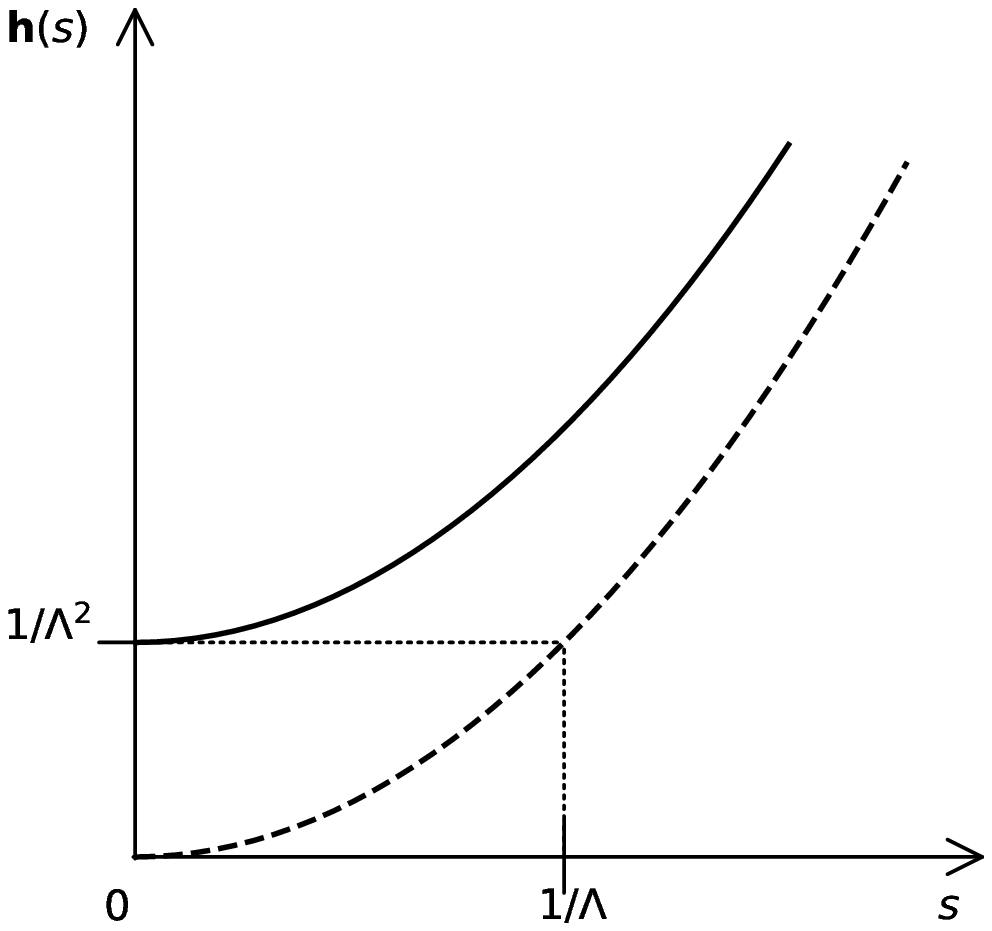}
\caption{In all figures a deformed function $\mathbf{h}(s)$ of the $s^2$ is depicted. The first one corresponds to $s^{2-\varepsilon}$. The others are related to (\ref{cuttuc1}) and (\ref{cuttuc2}), respectively. The dash-point line corresponds to $s^2$.}
\label{regs}
\end{figure}

\textbf{Naive approach: cutoff-1 and cutoff-2.} Now we move on to the second type of regularization. It preserves the dimension of the space ($d=4$) and can be introduced by a deformation of the interval $|x-y|^2$ in the exponential from formula (\ref{HK2}). There are a lot of ways to make this change, but we are interested in two approaches, that have appeared earlier in the papers \cite{w5,Ivanov-Kharuk-2020}. They can be defined according to the following formulae, see Figure \ref{regs},
\begin{align}\label{cuttuc1}
\mbox{Cutoff-1:}&\,\,\,\,\,\,|x|^2\to t^{\Lambda}_1(x)=
\begin{cases}
\,\,|x|^2, & |x|>1/\Lambda;\\
1/\Lambda^2, &|x|\leqslant 1/\Lambda,
\end{cases}
\\\label{cuttuc2}
\mbox{Cutoff-2:}&\,\,\,\,\,\,|x|^2\to t^{\Lambda}_2(x)=|x|^2+1/\Lambda^2,
\end{align}
where in the both cases $\Lambda$ is a dimension parameter of the regularization, such that the construction $|x|\Lambda$ is dimensionless. It is easy to verify that the limit $\Lambda\to+\infty$ removes the regularization.

In this case the regularized versions of the auxiliary functions (\ref{SD10}) have the form
\begin{equation}\label{cuttuc3}
R_0^{\Lambda,i}(x)=\frac{1}{4\pi^2t^{\Lambda}_i(x)},\,\,\,
R_1^{\Lambda,i}(x)=-\frac{\ln(t^{\Lambda}_i(x)\mu^2)}{16\pi^2},\,\,\,
R_2^{\Lambda,i}(x)=\frac{t^{\Lambda}_i(x)\big(\ln(t^{\Lambda}_i(x)\mu^2)-1\big)}{64\pi^2},
\end{equation}
where $i=1,2$, and $\mu$ is an auxiliary dimension parameter that takes a finite value.

Let us move on to the calculation. We start with the first type of regularization. In this case the functions $R_j^{\Lambda,1}(x)$, where $j=0,1,2$, does not satisfy relations (\ref{dmreg3}) and (\ref{dmreg4}). It means that we need to compute all integrals $\mathrm{I}_1$--$\mathrm{I}_{8}$ separately. Let us note that the region $|x|\leqslant1/\Lambda$ does not give a contribution to the integrals. Hence, we should consider only the region $|x|>1/\Lambda$. Then, using the basic formulae
\begin{equation}\label{cuttuc4}
\int_{1/\Lambda}^{1/\mu}\frac{dr}{r}=\ln(\Lambda/\mu)=L,\,\,\,
\int_{1/\Lambda}^{1/\mu}\frac{dr}{r}\ln(r\mu)=-\frac{1}{2}L^2,
\end{equation}
we get the results presented in the second column of the tables in Section \ref{sec:explicit}.

Answers for the second type of regularization can be obtained with some simplifications, because the objects $R_j^{\Lambda,2}(x)$, where $j=0,1,2$, satisfy relations (\ref{dmreg4}). Hence, we can express $\mathrm{I}_1$--$\mathrm{I}_{8}$ through some basic auxiliary integrals. They have the form
\begin{align}\label{cuttuc5}
\mathrm{I}_{\mathrm{aux}}^1&=\int_{0}^{1/\mu^2}ds\,\frac{s^2}{64\pi^6(s+1/\Lambda^2)^3}\stackrel{\mathrm{IR}}{=}\frac{1}{(4\pi)^4}\frac{8L}{\pi^2}
,\\\label{cuttuc6}
\mathrm{I}_{\mathrm{aux}}^2&=\int_{0}^{1/\mu^2}ds\,
\frac{-s\ln\big((s+1/\Lambda^2)\mu^2\big)}{\pi^2(4\pi)^4(s+1/\Lambda^2)^2}
\stackrel{\mathrm{IR}}{=}\frac{1}{(4\pi)^4}\frac{2(L^2-L)}{\pi^2},\\\label{cuttuc7}
\mathrm{I}_{\mathrm{aux}}^3&=\int_{0}^{1/\mu^2}ds\,\frac{s}{16\pi^4(s+1/\Lambda^2)^2}\stackrel{\mathrm{IR}}{=}\frac{1}{(4\pi)^4}32L.
\end{align}
Then, we have
\begin{equation}\label{cuttuc10}
\mathrm{I}_1\stackrel{\mathrm{IR}}{=}\frac{\pi^2c_2^2}{2^23}\mathrm{I}_{\mathrm{aux}}^1-
\frac{\pi^2c_2^2}{2}\mathrm{I}_{\mathrm{aux}}^2-
\frac{5c_2^2}{2^73}\mathrm{I}_{\mathrm{aux}}^3,\,\,\,
\mathrm{I}_8\stackrel{\mathrm{IR}}{=}\frac{\pi^2c_2^2}{2^43}\mathrm{I}_{\mathrm{aux}}^1+
\frac{\pi^2c_2^2}{2^73}\mathrm{I}_{\mathrm{aux}}^3,
\end{equation}
\begin{equation}\label{cuttuc9}
\mathrm{I}_3\stackrel{\mathrm{IR}}{=}\frac{\pi^2c_2^2}{2^4}\mathrm{I}_{\mathrm{aux}}^1-
\frac{\pi^2c_2^2}{2}\mathrm{I}_{\mathrm{aux}}^2,\,\,\,
\mathrm{I}_4\stackrel{\mathrm{IR}}{=}-\frac{\pi^2c_2^2}{2}\mathrm{I}_{\mathrm{aux}}^1,
\end{equation}
\begin{equation}\label{cuttuc8}
\mathrm{I}_5\stackrel{\mathrm{IR}}{=}-\frac{\pi^2c_2^2}{2^4}\mathrm{I}_{\mathrm{aux}}^1+
\frac{\pi^2c_2^2}{2}\mathrm{I}_{\mathrm{aux}}^2,\,\,\,
\mathrm{I}_6\stackrel{\mathrm{IR}}{=}-\frac{\pi^2c_2^2}{2^4}\mathrm{I}_{\mathrm{aux}}^1,\,\,\,
\mathrm{I}_7\stackrel{\mathrm{IR}}{=}\frac{\pi^2c_2^2}{2^5}\mathrm{I}_{\mathrm{aux}}^1.
\end{equation}

A contribution from $\mathrm{I}_2$ is a little bit different and can be obtained with the use of $\mathrm{I}_{\mathrm{aux}}^1$--$\mathrm{I}_{\mathrm{aux}}^3$ and the following equality
\begin{equation}\label{cuttuc11}
\int_{0}^{1/\mu^2}ds\,
\frac{s^3\ln\big((s+1/\Lambda^2)\mu^2\big)}{2(s+1/\Lambda^2)^4}
\stackrel{\mathrm{IR}}{=}-L^2+11L/6.
\end{equation}

Finally, after all calculations we get the third column in the tables in Section \ref{sec:explicit}.\\

\noindent\textbf{Cutoff-3, smoothed version of the cutoff-1.} In the previous section we have studied two types of cutoff regularization. Let us draw attention to the fact that no one satisfies reproducing equations (\ref{dmreg3}) in the form
\begin{equation}\label{smcut6}
-\partial_{x_\mu}\partial_{x^\mu}R_1(x)=R_0(x),\,\,\,
-\partial_{x_\mu}\partial_{x^\mu}R_2(x)=2R_1(x)-\frac{1}{16\pi^2}.
\end{equation}

So in this section we want to deform the cutoff-1 regularization in such way that the last equations would be satisfied. Moreover, we take the first function $R_0^{\Lambda,3}(x)=R_0^{\Lambda,1}(x)$ in the same form, see formulae (\ref{cuttuc1}) and (\ref{cuttuc3}). The next functions can be defined as follows
\begin{equation}
\label{smcut1}
R_1^{\Lambda,3}(x)=\frac{1}{4\pi^2}
\begin{cases}
-\frac{1}{4}\ln(|x|^2\mu^2)-\frac{1}{8}|x|^{-2}\Lambda^{-2},&|x|>1/\Lambda;\\
\,\,\,\,\,\,\,\,\,\,\,\,\,\,\,\,\,
\frac{1}{2}L-\frac{1}{8}|x|^2\Lambda^2,&|x|\leqslant1/\Lambda,
\end{cases}
\end{equation}
\begin{equation}
\label{smcut2}
4\pi^2R_2^{\Lambda,3}(x)=-\frac{\tilde{\alpha}L}{8\Lambda^2}+
\begin{cases}
\frac{1}{16}|x|^2\big(\ln(|x|^2\mu^2)-1\big)
+\frac{1}{16}\Lambda^{-2}\ln(|x|^2\Lambda^2)+\frac{1}{96}|x|^{-2}\Lambda^{-4}
+\frac{3}{32}\Lambda^{-2}
,&|x|>1/\Lambda;\\
\,\,\,\,\,\,\,\,\,\,\,\,\,\,\,\,\,\,\,\,\,\,\,\,\,\,\,\,\,\,\,\,\,\,\,\,\,\,\,\,\,\,\,
-\frac{1}{8}|x|^2L+\frac{1}{96}|x|^4\Lambda^2+\frac{1}{32}|x|^2
,&|x|\leqslant1/\Lambda,
\end{cases}
\end{equation}
where $\tilde{\alpha}$ is an auxiliary number from $\mathbb{R}$.

In addition to equalities (\ref{smcut1}) and (\ref{smcut2}), these functions also have the property of intermediate smoothness, which can be written as follows
\begin{equation}
\label{smcut7}
R_i^{\Lambda,3}(x)\Big|_{|x|=1/\Lambda-0}=
R_i^{\Lambda,3}(x)\Big|_{|x|=1/\Lambda+0},\,\,\,
\mbox{where}\,\,\,i=1,2.
\end{equation}

Additionally, we need to introduce two auxiliary functions $R_3^{\Lambda,3}(x)$ and $R_4^{\Lambda,3}(x)$, which solve the following equations
\begin{equation}
\label{smcut8}
-\frac{|x|^2}{16}R_0^{\Lambda,3}(x)+R_1^{\Lambda,3}(x)+\partial_{x_\mu}\partial_{x^\mu}
\bigg(
-\frac{|x|^2}{48}R_1^{\Lambda,3}(x)+\frac{5}{12}R_2^{\Lambda,3}(x)-
\frac{5}{2^93\pi^2}R_3^{\Lambda,3}(x)
\bigg)=0
\end{equation}
and
\begin{equation}
\label{smcut9}
-\frac{|x|^2}{16}R_0^{\Lambda,3}(x)+\partial_{x_\mu}\partial_{x^\mu}
\bigg(
-\frac{|x|^2}{48}R_1^{\Lambda,3}(x)-\frac{1}{12}R_2^{\Lambda,3}(x)+
\frac{1}{2^93\pi^2}R_4^{\Lambda,3}(x)
\bigg)=0,
\end{equation}
which are actually equalities from (\ref{green}), reformulated for (\ref{singord3}) and (\ref{singord4}). They have the form
\begin{equation}
\label{smcut10}
R_3^{\Lambda,3}(x)=|x|^2-\frac{8}{5}
\begin{cases}
\frac{1}{4}\Lambda^{-2}\ln(|x|^2\Lambda^2)+\frac{1}{6}|x|^{-2}\Lambda^{-4},&|x|>1/\Lambda;\\
\,\,\,\,
\frac{1}{24}\big(|x|^4\Lambda^2-\Lambda^{-2}\big)+\frac{1}{6}\Lambda^{-2},&|x|\leqslant1/\Lambda,
\end{cases}
\end{equation}
\begin{equation}
\label{smcut11}
R_4^{\Lambda,3}(x)=|x|^2+8
\begin{cases}
\frac{1}{4}\Lambda^{-2}\ln(|x|^2\Lambda^2)+\frac{1}{6}|x|^{-2}\Lambda^{-4},&|x|>1/\Lambda;\\
\,\,\,\,
\frac{1}{24}\big(|x|^4\Lambda^2-\Lambda^{-2}\big)+\frac{1}{6}\Lambda^{-2},&|x|\leqslant1/\Lambda.
\end{cases}
\end{equation}

Now we are ready to proceed the calculations. Following the general idea we need to compute integrals (\ref{aped17})--(\ref{aped16}) with the use of new formulae. Fortunately, we can do some simplifications. Indeed, we can note that the integrals $\mathrm{I}_1$ and $\mathrm{I}_4$--$\mathrm{I}_9$ have the same singularities as in the case of the cutoff-1 regularization. Hence, we need to compute only three objects: $\mathrm{I}_2$, $\mathrm{I}_3$, $\mathrm{I}_{10}$. 

All results are presented in the two tables in Section \ref{sec:explicit}.

\subsection{Tables of form factors}
\label{sec:explicit}
In the section we present our calculations in the form of two tables. In the first one we give the singularities of integrals (\ref{aped17})--(\ref{aped16}) for different types of regularization: dimensional one from Section \ref{sec:exam:dr}, cutoff-1, cutoff-2, and cutoff-3 from Section \ref{sec:exam:cut}.

\begin{center}
	\begin{tabular}{c c c c c}
		&$\mbox{\footnotesize{\textbf{Dim. reg.}}}$&
		$\mbox{\footnotesize{\textbf{Cutoff-1 reg.}}}$&
		$\mbox{\footnotesize{\textbf{Cutoff-2 reg.}}}$&
		$\mbox{\footnotesize{\textbf{Cutoff-3 reg.}}}$\\
		$\mbox{\footnotesize{Integral}}$&
		$\frac{(4\pi)^4\mu^{2\varepsilon}}{c_2^2}\Big(\substack{\mbox{\footnotesize{IR }}\\\mbox{\footnotesize{part}}}\Big)$&
		$\frac{(4\pi)^4}{c_2^2}\Big(\substack{\mbox{\footnotesize{IR }}\\\mbox{\footnotesize{part}}}\Big)$&
		$\frac{(4\pi)^4}{c_2^2}\Big(\substack{\mbox{\footnotesize{IR }}\\\mbox{\footnotesize{part}}}\Big)$&
		$\frac{(4\pi)^4}{c_2^2}\Big(\substack{\mbox{\footnotesize{IR }}\\\mbox{\footnotesize{part}}}\Big)$\\
		$\mathrm{I}_1$
		&$-1/\varepsilon^2-5/8\varepsilon$
		&$-L^2-L/4$&$-L^2+5L/4$& $-L^2-L/4$ \\
		$\mathrm{I}_2$
		&$1/\varepsilon^2+5/8\varepsilon$
		&$L^2+5L/4$&$L^2-11L/36$& $L^2+5L(1/4+\tilde{\alpha}/6)$ \\
		$\mathrm{I}_3$
		&$-1/\varepsilon^2-1/4\varepsilon$
		&$-L^2$&$-L^2+3L/2$& $-L^2+L/2$\\
		$\mathrm{I}_4$
		&$-2/\varepsilon$
		&$-4L$&$-4L$ & $-4L$\\
		$\mathrm{I}_5$
		&$1/\varepsilon^2+1/4\varepsilon$
		&$L^2$&$L^2-3L/2$& $L^2$\\
		$\mathrm{I}_6$
		&$-1/4\varepsilon$
		&$-L/2$&$-L/2$& $-L/2$\\
		$\mathrm{I}_7$
		&$1/8\varepsilon$
		&$L/4$&$L/4$& $L/4$\\
		$\mathrm{I}_8$
		&$1/8\varepsilon$
		&$L/4$&$L/4$& $L/4$\\
		$\mathrm{I}_9$
		&$4/\varepsilon^2$
		&$4L^2$&$4L^2$& $4L^2$\\
		$\mathrm{I}_{10}$
		&$0$
		&$-2L$&$-2L$&$-2\tilde{\alpha}L$
	\end{tabular}
\end{center}

In the second table we present several linear combinations of the integrals, computed above, such as contribution (\ref{twoloop112}) to the pure effective action (\ref{eq46}) and its separate parts (\ref{twoloop103})--(\ref{twoloop107}), (\ref{twoloop109}), and (\ref{twoloop111}). Also, we study additional counterterm (\ref{qemm14}) from Section \ref{sec:twol:quant} to compare the answer for the dimensional regularization.

\begin{center}
	\begin{tabular}{c c c c c}
		&$\mbox{\footnotesize{\textbf{Dim. reg.}}}$&
		$\mbox{\footnotesize{\textbf{Cutoff-1 reg.}}}$&
		$\mbox{\footnotesize{\textbf{Cutoff-2 reg.}}}$&
		$\mbox{\footnotesize{\textbf{Cutoff-3 reg.}}}$\\
		$\mbox{\footnotesize{Contribution}}$&
		$\frac{(4\pi)^4\mu^{2\varepsilon}}{g^2c_2^2}\Big(\substack{\mbox{\footnotesize{IR }}\\\mbox{\footnotesize{part}}}\Big)$&
		$\frac{(4\pi)^4}{g^2c_2^2}\Big(\substack{\mbox{\footnotesize{IR }}\\\mbox{\footnotesize{part}}}\Big)$&
		$\frac{(4\pi)^4}{g^2c_2^2}\Big(\substack{\mbox{\footnotesize{IR }}\\\mbox{\footnotesize{part}}}\Big)$&
		$\frac{(4\pi)^4}{g^2c_2^2}\Big(\substack{\mbox{\footnotesize{IR }}\\\mbox{\footnotesize{part}}}\Big)$\\
		$\mathcal{J}_1$
		&$-4/\varepsilon^2-5/2\varepsilon$
		&$-4L^2$&$-4L^2+53L/9$& $-4L^2+5\tilde{\alpha}L/3$ \\
		$\mathcal{J}_2$
		&$8/\varepsilon^2+3/\varepsilon$
		&$8L^2$&$8L^2-106L/9$& $8L^2-10\tilde{\alpha}L/3$ \\
		$\mathcal{J}_3$
		&$2/\varepsilon^2+1/\varepsilon$
		&$2L^2-L$&$2L^2-35L/9$& $2L^2-L(2+5\tilde{\alpha}/3)$\\
		$\mathcal{J}_4$
		&$1/8\varepsilon$
		&$-L/2$&$-17L/36$& $-L-5\tilde{\alpha}L/12$\\
		$-\sum_{n=1}^4\mathcal{J}_n$
		&$-6/\varepsilon^2-13/8\varepsilon$
		&$-6L^2+3L/2$&$-6L^2+41L/4$& $-6L^2+3L(1+3\tilde{\alpha}/4)$\\
		$-\mathcal{J}_5$
		&$-3/8\varepsilon$
		&$-L$&$-35L/36$& $-L(1+5\tilde{\alpha}/12)$\\
		$-\mathcal{J}_6$
		&$6/\varepsilon^2$
		&$6L^2-5L$&$6L^2-5L$& $6L^2-5\tilde{\alpha}L$\\
		$-\sum_{n=1}^6\mathcal{J}_n$
		&$-2/\varepsilon$
		&$-9L/2$&$77L/18$& $L(2-5\tilde{\alpha}/3)$\\
		$-\mathcal{J}_7$&$-5/6\varepsilon$&$\mbox{\footnotesize{not exist}}$&$-25L/36$&$0$\\
		$-\sum_{n=1}^7\mathcal{J}_n$&$-17/6\varepsilon$&---&$43L/12$&$L(2-5\tilde{\alpha}/3)$
	\end{tabular}
\end{center}

Let us comment the last results. First of all, we note that our formula (\ref{twoloop112}) reproduces the correct results for the second loop in the case of dimensional regularization, see \cite{12}. Thus, we have checked it.

Secondly, we draw attention to the fact, that the counterterm in the case of cutoff-1 can not be calculated, because the regularization after the first derivative loses the smoothness near the diagonal. Of course, it is possible to compute it by using the determinant of the operator \cite{Shore}, but it is not the main aim of our paper.

At the same time we have obtained the same value for the divergent part of the pure effective action (\ref{eq46}) in the case of cutoff-1, as it was calculated in \cite{Ivanov-Kharuk-2020}. Additionally, we have got the results for two supplemental regularizations, one of which depends on the auxiliary parameter that can be chosen based on additional physical considerations.

\subsection{Shift of a special type}
\label{sec:shift}
In this section we are going to present the fourth type of cutoff regularization, which is based on a shift of special type of the cutoff-3, see \cite{Ivanov-Kharuk-2020}. Indeed, we can deform the function $R_0^{\Lambda,3}(x)$ in the region $|x|\leqslant1/\Lambda$ in the following form
\begin{equation}\label{smcut14}
R_0^{\Lambda,3}(x)\to
R_0^{\Lambda,4}(x)=R_0^{\Lambda,3}(x)+\tilde{R}_0^{\Lambda}(x),\,\,\,
\tilde{R}_0^{\Lambda}(x)=\frac{1}{4\pi^2}
\begin{cases}
\,\,\,\,\,\,\,\,\,\,\,\,0,&|x|>1/\Lambda;\\
\Lambda^2f_0\big(\Lambda^2|x|^2\big),&|x|\leqslant1/\Lambda,
\end{cases}
\end{equation}
where the auxiliary function has the following properties: $f_0(\cdot)\in C^{\infty}\big([0,1],\mathbb{R}\big)$, $\partial_{x_\mu}\partial_{x^\mu}\Lambda^2f_0\big(\Lambda^2|x|^2\big)\to0$ in the sense of generalized functions for $\Lambda\to+\infty$, and $f_0(1)=0$.

Then, according to the general idea, described above, we need to find such $\tilde{R}_i^{\Lambda}(x)$, $i=1,2$, that equalities (\ref{smcut6}) would be satisfied for $R_i^{\Lambda,3}(x)\to R_i^{\Lambda,3}(x)+\tilde{R}_i^{\Lambda}(x)$. This leads to the relations
\begin{equation}\label{smcut12}
-\partial_{x_\mu}\partial_{x^\mu}\tilde{R}_1^{\Lambda}(x)=\tilde{R}_0^{\Lambda}(x),\,\,\,
-\partial_{x_\mu}\partial_{x^\mu}\tilde{R}_2^{\Lambda}(x)=2\tilde{R}_1^{\Lambda}(x).
\end{equation}

They can be integrated in a very simple way. Firstly, let us note that the ordinary Laplace operator $\partial_{x_\mu}\partial_{x^\mu}$ has the following form $r^{-3}\partial_rr^3\partial_r$, where $r=|x|$, in the polar coordinates, in the case of applying to the spherically-symmetric functions. Secondly, let us define the following operation
\begin{equation}\label{smcut15}
\psi:C^{\infty}\big([0,1],\mathbb{R}\big)\to C^{\infty}\big([0,1],\mathbb{R}\big),
\end{equation}
which acts according to the formula
\begin{equation}\label{smcut16}
\psi(f)(\tau)=-\frac{1}{4}\int_0^\tau dt\,t^{-2}\int_0^t ds\,sf(s),\,\,\,\mbox{for all}
\,\,\,f\in C^{\infty}\big([0,1],\mathbb{R}\big)\,\,\,\mbox{and}\,\,\,\tau\in[0,1].
\end{equation}

Further, we introduce some auxiliary objects
\begin{align}\label{smcut17}
f_1&=\psi(f_0)\in C^{\infty}\big([0,1],\mathbb{R}\big),\\
a(f_0^{\phantom{\prime}})&=\frac{1}{4}\int_0^1 ds\,sf_0^{\phantom{\prime}}(s)=-f_1^\prime(1)\in\mathbb{R},\\
b(f_0^{\phantom{\prime}})&=\frac{1}{4}\int_0^1 ds\,f_0^{\phantom{\prime}}(s)=-f_1^\prime(1)-
f_1^{\phantom{\prime}}(1)\in\mathbb{R}.
\end{align}

After all the preparations, we can write out the answer in the form

\begin{equation}
\label{smcut13}
\tilde{R}_1^{\Lambda}(x)=\frac{1}{4\pi^2}
\begin{cases}
\,\,\,\,\,a(f_0)|x|^{-2}\Lambda^{-2},&|x|>1/\Lambda;\\
f_1\big(|x|^2\Lambda^2\big)+b(f_0),&|x|\leqslant1/\Lambda,
\end{cases}
\end{equation}
\begin{equation}
\label{smcut18}
\tilde{R}_2^{\Lambda}(x)=\frac{1}{4\pi^2}
\begin{cases}
\,\,\,
-\frac{1}{2}\Lambda^{-2}a(f_0)\ln\big(|x|^{2}\Lambda^{2}\big)
+|x|^{-2}\Lambda^{-4}\big(-\frac{1}{2}a(f_0)+2a(f_1)+\frac{1}{4}b(f_0)\big),&|x|>1/\Lambda;\\
2\Lambda^{-2}\psi(f_1)\big(|x|^2\Lambda^2\big)
-\frac{1}{4}|x|^2b(f_0)
+\Lambda^{-2}\big(-\frac{1}{2}a(f_0)+2b(f_1)+
\frac{1}{2}b(f_0)\big)
,&|x|\leqslant1/\Lambda,
\end{cases}
\end{equation}
where the continuity properties of the first derivative were used.
In the same way we can reformulate and solve equations (\ref{smcut10}) and (\ref{smcut11}).
So we get for $i=3,4$
\begin{equation}
\label{smcut19}
\frac{\rho_i}{2^93\pi^2}\tilde{R}_i^{\Lambda}(x)=-\frac{1}{12}\tilde{R}_2^{\Lambda}(x)-
\frac{|x|^2}{48}\tilde{R}_1^{\Lambda}(x)
+\frac{\Lambda^{-2}}{64\pi^2}
\begin{cases}
\,\,\,\,\,\,\,\,\,a(\hat{f}_0)|x|^{-2}\Lambda^{-2},&|x|>1/\Lambda;\\
\psi(\hat{f}_0)\big(|x|^2\Lambda^2\big)+b(\hat{f}_0),&|x|\leqslant1/\Lambda,
\end{cases}
\end{equation}
where $\hat{f}_0(s)=sf_0(s)$, $\rho_3=5$, and $\rho_4=-1$.

Now we are ready to calculate the integrals (\ref{aped17})--(\ref{aped16}). Firstly, we note that it is convenient to use for computing the results for the cutoff-3 case from the tables in Section \ref{sec:explicit}. For example, the integrals $\mathrm{I}_4$, $\mathrm{I}_6$, $\mathrm{I}_7$, and $\mathrm{I}_8$ are not violated. So they equal
\begin{equation}
\label{smcut20}
\mathrm{I}_4\stackrel{\mathrm{IR}}{=}
\frac{c_2^2}{(4\pi)^4}\big(-4L\big),\,\,\,
\mathrm{I}_6\stackrel{\mathrm{IR}}{=}
\frac{c_2^2}{(4\pi)^4}\big(-L/2\big),\,\,\,
\mathrm{I}_7\stackrel{\mathrm{IR}}{=}
\frac{c_2^2}{(4\pi)^4}\big(L/4\big),\,\,\,
\mathrm{I}_8\stackrel{\mathrm{IR}}{=}
\frac{c_2^2}{(4\pi)^4}\big(L/4\big).
\end{equation}
The next group of integrals has additional terms. Then, using (\ref{smcut13}) and (\ref{smcut18}) we get
\begin{align}
\label{smcut21}
\mathrm{I}_1\stackrel{\mathrm{IR}}{=}&
\frac{c_2^2}{(4\pi)^4}\bigg(-L^2-L/4-8La(f_0^{\phantom{2}})-L\int_0^1ds\,sf_0^2(s)\bigg),\\\label{smcut22}
\mathrm{I}_2\stackrel{\mathrm{IR}}{=}&
\frac{c_2^2}{(4\pi)^4}\bigg(L^2+5L(1/4+\tilde{\alpha}/6)+L\int_0^1ds\,s^3\big(f_0^\prime(s)\big)^2\bigg),\\\label{smcut23}
\mathrm{I}_3\stackrel{\mathrm{IR}}{=}&
\frac{c_2^2}{(4\pi)^4}\bigg(-L^2+L/2-8La(f_0^{\phantom{2}})-L\int_0^1ds\,sf_0^2(s)\bigg),\\\label{smcut24}
\mathrm{I}_5\stackrel{\mathrm{IR}}{=}&
\frac{c_2^2}{(4\pi)^4}\bigg(L^2+8La(f_0)+L\int_0^1ds\,sf_0^2(s)\bigg).
\end{align}
Further, the diagonal parts are equal to
\begin{equation}
\label{smcut25}
\mathrm{I}_9\stackrel{\mathrm{IR}}{=}
\frac{c_2^2}{(4\pi)^4}\big(4L^2+16Lb(f_0)\big),\,\,\,
\mathrm{I}_{10}\stackrel{\mathrm{IR}}{=}
\frac{c_2^2}{(4\pi)^4}\big(-2\tilde{\alpha}L(1+f_0(0))\big).
\end{equation}

Hence, after all summations we get
\begin{equation}
\label{smcut26}
-\sum_{n=1}^6\mathcal{J}_n\stackrel{\mathrm{IR}}{=}\frac{g^2c_2^2}{(4\pi)^4}
L\bigg(2-5\tilde{\alpha}\big(1/3+f_0(0)\big)-80
a(f_0^{\phantom{2}})+24b(f_0^{\phantom{2}})-10\int_0^1ds\,sf_0^2(s)+4\int_0^1ds\,s^3\big(f_0^\prime(s)\big)^2
\bigg).
\end{equation}

For example, if we take $f_0(s)=1-s$, then we get
\begin{equation}
\label{smcut27}
a(f_0^{\phantom{2}})=\frac{1}{24}
,\,\,\,
b(f_0^{\phantom{2}})=\frac{1}{8}
,\,\,\,
\int_0^1ds\,sf_0^2(s)=\frac{1}{12}
,\,\,\,
\int_0^1ds\,s^3\big(f_0^\prime(s)\big)^2=\frac{1}{4},
\end{equation}
and
\begin{equation}
\label{smcut28}
-\sum_{n=1}^6\mathcal{J}_n\stackrel{\mathrm{IR}}{=}
\frac{g^2c_2^2}{(4\pi)^4}L
(11-40\tilde{\alpha})/6.
\end{equation}
The last example describes the cutoff regularization that preserves the continuity of the first and the second derivatives of the function $R_0\big|_{\scriptsize{\mbox{IR-reg.}}}$. As we see, there is one additional free parameter.

\section{Conclusion}
\label{sec:concl}
In the present work we have derived new formula (\ref{effact1}) for the two-loop contribution to the pure effective action (\ref{eq46}). This formula is universal and can be used for any type of the regularization that does not deform the Seleey--DeWitt coefficients. Actually, the answer depends on the three functions (\ref{SD10}) from the heat kernel expansion and their deformation in the process of regularization, see, for example, (\ref{dmreg1}), (\ref{cuttuc3}), (\ref{smcut1}), and (\ref{smcut14}). 

To verify the correctness of the obtained formula (\ref{twoloop112}), we performed calculations for several types of regularization, such as dimensional one and cutoff one in several forms, see the tables from Section \ref{sec:explicit}. All the results are consistent with those obtained earlier in other papers, see \cite{Ivanov-Kharuk-2020,12}. Moreover, we have shown that all regularizations do not lead to double-logarithmic ($L^2$) and non-logarithmic ($\Lambda$ and $\Lambda^2$) singularities. At the same time we need to draw attention to the fact that the singularities from $\Gamma_4$-term differ from other ones, because they depend only on the value of regularized functions (\ref{SD10}) at zero, while other divergencies depend on a behaviour in some neighborhood. In some sense they have a different nature that can be studied in further.

Also, we should note that in the case of general cutoff regularization, we have some auxiliary parameters. We believe that they will be concretized after satisfying additional physical requirements. As an example of such conditions we can give the gauge invariance. Some useful remarks on its restoring can be found in the papers \cite{w3,w4,w9,w10}. We hope, that such conditions would give a relation between singularities of two types (at zero and near zero), mentioned in the previous paragraph.

Additionally, we need to note that the consideration of a regularization that transforms the Seeley--DeWitt coefficients as well is also possible. In this case we should use formulae (\ref{aped1})--(\ref{aped7}) and (\ref{aped8}) from the proof instead of (\ref{aped17})--(\ref{aped16}). The detailed description of such types of regularization is not included in the present work and will appear later.

\paragraph{Acknowledgements.}
This research is fully supported by the grant in the form of subsidies from the Federal budget for state support of creation and development world-class research centers, including international mathematical centers and world-class research centers that perform research and development on the priorities of scientific and technological development. The agreement is between MES and PDMI RAS from \textquotedblleft8\textquotedblright\,November 2019 \textnumero\ 075-15-2019-1620.

\end{document}